\documentclass[12pt]{elsarticle}
\usepackage{graphicx}
\usepackage{amsmath}
\usepackage{bm}

\journal{and accepted by Journal of Engineering Mathematics 2020}
\def\bba{\begin{align}}
\def\bea{\end{align}}
\def\ba{\begin{align*}}
\def\ea{\end{align*}}

\def\ee{\end{eqnarray*}}
\def\be{\begin{eqnarray*}}
\def\bee{\end{eqnarray}}
\def\bbe{\begin{eqnarray}}

  \def\R{\mathrm{Re}}
   
   \def\i{\mathrm i}
   \def\p{\partial}
   \def\e{\mathrm e}
   \def\H{\mathcal H}

   \def\n{\bm n}
    \def\e{\mathrm e}

    \def\p{\partial}
    \def\q{\bm q} 

    \def\i{\mathrm i}

\def\d{\mathrm{d}}
\def\panel{\mathrm{panel}}

\def\D{D}

\def\pp{{\bm p}}
\title{New formulation of the finite depth free surface Green function}

\author{Zhi-Min Chen}
\address{School  of Mathematics and Statistics, Shenzhen University,  Shenzhen 518060,  China}
\date{}

\begin{document}
\begin{abstract}
 For a pulsating free surface source in a three-dimensional finite depth fluid domain, the  Green function  of the source presented by John [F. John,  On the motion of floating bodies II. Simple harmonic motions, Communs. Pure Appl. Math. 3 (1950) 45-101] is  superposed as the Rankine source potential, an image source potential and a wave integral in the infinite domain $(0, \infty)$. When the source point together with a field point is on the free surface, John's  integral and  its gradient are not convergent since  the integration  $\int^\infty_\kappa$ of the corresponding integrands does  not  tend to zero in a uniform manner  as   $\kappa$ tends to $\infty$. Thus evaluation of the Green function is not based on direct integration of the wave integral but is obtained by  approximation expansions in earlier investigations. In the present study, five images of the source with respect to  the free surface mirror and the water bed mirror in relation to the image method are employed to reformulate the wave integral. Therefore the free surface Green function of the source is decomposed into the Rankine  potential, the five image source potentials and a new wave integral, of which the integrand is  approximated by a smooth and rapidly decaying function. The gradient of the Green function is further formulated so that the same integration stability with the wave integral is demonstrated. The significance of the present research is that the   improved  wave integration  of the Green function and its gradient becomes convergent.   Therefore evaluation of the Green function is obtained through the  integration of the integrand in a straightforward manner.  The application of the scheme  to a floating body or a submerged body motion in regular waves  shows that the approximation is sufficiently accurate to compute linear wave loads in practice.
\end{abstract}

\maketitle


\section{Introduction}

The understanding  of wave induced forces  resulting from a wave-body motion  is  fundamental  in  hydrodynamics.  In the linear potential flow theory, the velocity potential of the fluid motion problem is a harmonic function  and  can be  represented as a solution of body boundary integral equation involving   free surface Green function. The integral equation can be solved numerically  by combining  a boundary element method  for the numerical integration of the  free surface Green function or free surface sources distributed on wetted body surface (see, for example, Frank \cite{Frank1967}, Lee and Sclavounos \cite{Lee1989}, Lee and Newman \cite{LeeNew} for the infinite water depth).  Thus it is fundamental for the  evaluation of  the free surface Green function (see, for example, the  successful investigations given by Chakrabarti \cite{Chak}, Liang {\it et al.} \cite{LiangWuNob18}, Newman   \cite{N1985}, Noblesse \cite{Nob82}, Ponizy {\it et al.} \cite{PonBob94}, Telste and Noblesse \cite{TelNob86}, Wu {\it et al.} \cite{WuNob17} for the  infinite  depth case and John \cite{Jon}, Linton \cite{Linton1999}, Liu {\it et al.} \cite{Liu},    Newman \cite{N1985}, Pidcock \cite{Pid}  for the  finite  depth case).

For a radial symmetric body
undergoing an oscillatory motion in a fluid of   infinite water depth, its linear analytic solution can be approximated by  a single free surface source rather than the boundary integral of free surface sources continuously distributed on the body surface.  For a heaving or surging hemisphere, the velocity potential  solution  is   decomposed into a  free surface source located at the centre of the sphere and a wave-free potential, which is expanded in a series of Legendre polynomials and sinusoidal functions (see, for example, Havelock \cite{Have1955}, Hulme \cite{Hu1982} and Ursell \cite{U1949}).  The unknown source strength and expansion coefficients  are determined by the boundary condition of the velocity potential on the hemisphere. This method also applies to the wave resistance problem (see Farell \cite{Fa}) of a travelling  spheroid  and an oscillatory  submerged sphere in  a fluid   (see Chatjigeorgiou \cite{Cha2013}, Wang \cite{W1986},  Wu and Eatock Taylor \cite{WuTaylor87}  of  infinite water depth and Linton
\cite{Linton1991} of  finite water depth). Satisfactory numerical solutions can also obtained from  varieties of Rankine simple source methods on  the wave body motion problem (see, for example, Cao {\it et al.} \cite{Cao},  Dawson \cite{Dawson}, Feng {\it et al.} \cite{FengChen1,FengChen2}, Mantzaris \cite{Man},  Yeung \cite{Y1981}) by using  the dynamic and kinematic free surface boundary conditions rather than the Green function theory.

 In the study of the pulsating free surface   Green function, the author \cite{Chen2019} showed the singular wave integral of the Green function being  approximated by a regular wave integral and the integration can be evaluated  directly by using Bessel functions.  The present study  is a continuation of \cite{Chen2019}  to  an oscillatory motion in a fluid of   finite water depth.

\begin{figure}[h]
\centering 
\includegraphics[width=.8\textwidth,height=.4\textwidth]{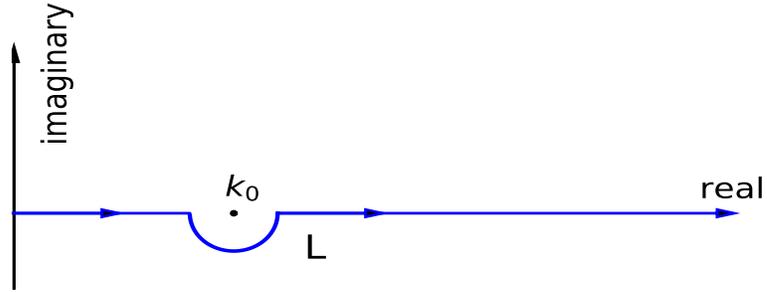}

\caption{ Profile of the integration path $L$  in  (\ref{G1}) passing beneath the positive root $k=k_0$.  }
\label{f1}
\end{figure}

Consider a pulsating free surface source $\pp=(\xi,\eta,\zeta)$, in a finite depth  fluid domain $-h<z< 0$,  undergoing an oscillatory motion of a frequency $\omega$.
The velocity potential of the linear wave motion  in the frequency domain  with respect to a field point $\q=(x,y,z)$ is expressed as
\be \mathcal G = \frac1{4\pi}\R( G e^{-\i\omega t}),\,\,\, \i=\sqrt{-1},\ee
where $G$ is  a complex function expressed  as
\be G=\frac1 {r} +\frac1{r_0}+ G_1
\ee
with $G_1$ a harmonic function in the fluid domain and
\be  r= \sqrt{(x-\xi)^2+(y-\eta)^2+(z-\zeta)^2} \mbox{ and } \,\,r_0= \sqrt{(x-\xi)^2+(y-\eta)^2+(z+\zeta+2h)^2}.
\ee
The function $G$ is known as a finite depth free surface Green function or  the fundamental solution of the Laplacian equation in the fluid domain associated with the free surface boundary condition and water bed boundary condition. As the Rankine source potential $\frac1 r$ is the fundamental solution of the Laplacian equation in the whole three-dimensional domain, the function $G$ is determined by the following the boundary value problem:
\bbe
&&\frac{\partial^2 G_1}{\partial x^2} +\frac{\partial^2 G_1}{\partial y^2} +\frac{\partial^2 G_1}{\partial z^2} =0,\,\,\, -h< z<0,\label{con1}
\\
&&\left.\frac{\partial G}{\partial z} -\nu G\right|_{z=0} =0, \label{con2}
\\
&&\left.\frac{\partial  G}{\partial z}\right|_{z=-h} =0,\ \ \ \ \label{con3}
\\
&&\lim_{R\to \infty} \sqrt{R}\left(\frac{\partial  G}{\partial R} -\i \nu G\right)=0\label{con4}
\bee
for  $R=|(x,y)-(\xi,\eta)|$ the horizontal distance, $g$ the gravitational acceleration and $\nu$ the dimensional wave number $ \omega^2/g$.
The Green function was obtained by John \cite[Eq. (A9)]{Jon} as
\bbe
 G&=&\frac1 r+\frac1 {r_0}+\int_L\frac{2(\nu +k)\e^{-kh}\cosh k(\zeta+h)\cosh k(z+h)}{k\sinh kh-\nu \cosh kh }J_0(kR)dk\label{G1}
\bee
for  $J_0$ the Bessel function of the first kind. The wave integral pass $L$  is illustrated in Figure \ref{f1}.
The harmonic function  $G_1$ is an irregular wave integral due to the occurrence  of the  pole $k_0>0$,  the positive root of the dispersion relation
\bbe k\tanh kh =\nu.\label{dis}
\bee
 A well known evaluation expansion of the Green function is given by John \cite[pp. 93-95]{Jon} as
\bbe
G&=&2\pi \frac{\nu^2-k_0^2}{hk_0^2-h\nu^2+\nu} \cosh k_0(z+h) \cosh k_0(\zeta+h)(  Y_0(k_0R) -\i J_0(k_0R))\nonumber 
\\
&&+4\sum_{n=1}^\infty\frac{k_n^2+\nu^2}{h k_n^2+h \nu^2-\nu} \cos k_n(z+h)\cos k_n(\zeta+h) K_0(k_n R)\label{GG3}
\\
&\approx &2\pi \frac{\nu^2-k_0^2}{hk_0^2-h\nu^2+\nu} \cosh k_0(z+h) \cosh k_0(\zeta+h)(  Y_0(k_0R) -\i J_0(k_0R))\nonumber 
\\
&&+4\sum_{n=1}^{N_J}\frac{k_n^2+\nu^2}{h k_n^2+h \nu^2-\nu} \cos k_n(z+h)\cos k_n(\zeta+h) K_0(k_n R)\label{GG33}
\bee
for $N_J$ a suitable positive integer,  $Y_0$ the Bessel function of the second kind,  $K_0$ the modified Bessel function of the second kind and  $\i k_n$  the  roots of the dispersion relation (\ref{dis})
satisfying
\be \pi (n-\frac12) \leq k_n h \le \pi n,\,\,\, n\ge 1.
\ee For further understanding of this expansion, one may refer to Wehausen and Laitone \cite[Eq. (13.19)]{WehausenLaitone} and  Newman \cite{N1985}.
Eq. (\ref{GG3}) exhibits a simple form of the Green function evaluation, which however is not a harmonic function due to the inclusion of the  Rankine source singular potential $1/r$. Thus  it is numerically inefficient within  finite boundary elements discretisation in wave body motions. What is more, as noticed by Newman \cite[p. 64]{N1985},
the series (\ref{GG3}) is
practically useless for small values of $R/h$, since each summand contains a logarithmic
singularity when $R/h = 0$.
Successful developments on John's evaluation series have been obtained by many authors  (see, for example,   Newman \cite{N1985}, Liu {\it et al.} \cite{Liu}, Linton \cite{Linton1999}, Pidcock \cite{Pid}).

The purpose of the present study is to  approximate  the three-dimensional finite depth free surface Green function through direct integration together with its application  to  wave body motions.

The integrand of the wave integral (\ref{G1}) has the asymptotic behaviour \cite{Abr}  (as $k\to \infty$):
\bbe \frac{2(\nu +k)\e^{-kh}\cosh k(\zeta+h)\cosh k(z+h)}{k\sinh kh-\nu \cosh kh }J_0(kR)\sim \e^{k(z+\zeta)}\frac{2\sqrt{2}\cos (kR-\frac\pi4)}{
\sqrt{\pi k R}}.\label{new11}
\bee
When both the field point $\q$ and the source point $\pp$ tend to the free surface $z=\zeta=0$, the wave integral (\ref{G1}) is not stable but oscillates with unbounded amplitude.

What is more, the wave integral stability for the gradient of the Green function  is even  worse. For example, the integrand of the horizontal derivative
 \bbe
 \p_R G&=&\p_R(\frac1 r+\frac1 {r_0})-\int_L\frac{2(\nu +k)\e^{-kh}\cosh k(\zeta+h)\cosh k(z+h)}{k\sinh kh-\nu \cosh kh }kJ_1(kR)dk\label{new22}
\bee
has the asymptotic behaviour ( as $k\to \infty$)
\bbe \frac{2(\nu +k)\e^{-kh}\cosh k(\zeta+h)\cosh k(z+h)}{k\sinh kh-\nu \cosh kh }kJ_1(kR)\sim \frac{2\sqrt{2k}\cos (kR-\frac{3\pi}4)}{
\sqrt{\pi  R}}\label{new33}
\bee
on the free surface $z=\zeta=0$.

Thus it is beneficial   to provide a stable formulation of the Green function. To do so, we  introduce a new formulation of the Green function   expressed as
\bbe
G &=&\frac1r + \frac1{r_0}+\frac1{r_1}+\frac1{r_2}+\frac1{r_3}+\frac1{r_4}+K\label{myG2}
\bee
with the wave integral as the limit of smooth function integrals
\bbe
K=\lim_{\mu\to 0+}  \int^\infty_0\frac{[2\nu \!+\!(k\!+\!\nu)e^{-2kh}][\e^{k(z\!+\!\zeta)}\!+\!\e^{ k(z-\zeta-2h)}\!+\!\e^{k(-z\!+\!\zeta-2h)}\!+\!\e^{ -k(z\!+\!\zeta\!+\!4h)}]J_0(kR)dk}{(1\!+\!e^{-2kh})(k\tanh kh-\nu-\i\mu ) }  \label{G2}
\bee
and
\be &&r_1= \sqrt{R^2+(z+\zeta)^2},\,\,\,r_2=\sqrt{R^2+(z-\zeta+2h)^2},\,\,\,
\\&& r_3=\sqrt{R^2+(\zeta-z+2h)^2},\,\, r_4=\sqrt{R^2+(z+\zeta+4h)^2}.
\ee
Similar, new  formulation for the  gradient Green function  $\nabla G$, which has the same wave integral stability with the new formulation of $G$,  is also obtained. For example, we have the horizontal derivative formulation
\begin{align} \p_R& G = \partial_R \left(\frac1{r}+\frac1{r_0}+\frac1{r_1}+\frac1{r_2}+\frac1{r_3}+\frac1{r_4}\right)\nonumber\\
 & \nonumber -\!\!\!\!\lim_{\mu\to 0+}\int^\infty_0\!\!\frac{(2\nu^2\!+\!(k\!+\!2\nu)(k\!+\!\nu)e^{-2kh}) (\e^{k(z+\zeta)}\!+\!\e^{ k(z-\zeta-2h)}\!+\!\e^{k(\zeta-z-2h)}\!+\!\e^{ -k(z+\zeta+4h)})  J_1(kR)dk}{(1\!+\!e^{-2kh})(k\tanh kh\!-\!\nu\!-\!\i\mu ) }
\\ \label{new44}
&-\left( \frac{2\nu R}{r_1(r_1\!+\!|z\!+\!\zeta|)}\!+\!\frac{2\nu R}{r_2(r_2\!+\!|z\!-\!\zeta\!-\!2h|)}\!+\!\frac{2\nu R}{r_3(r_3\!+\!|\zeta\!-\!z\!-\!2h|)}\!+\!\frac{2\nu R}{r_4(r_4\!+\!|z\!+\!\zeta\!+\!4h|)}\right).
\end{align}

The parameter $\mu$ is known as the Rayleigh artificial viscosity, which was used by Havelock \cite{Hav28,Hav32} by employing the limit $\mu\to 0$ to show the uniqueness of  the infinite water depth  Green function of a free surface source advancing at a uniform speed. In the present study, however, we use the regular wave integral with $\mu>0$  to cancel the singularity  around the pole $k=k_0$ in the direct  integration scheme rather than  take  the limit $\mu\to0$ to describe the troublesome of the singularity in earlier investigations.  For free surface wave damping under the effect of viscosity, one may refer to Lazauskas\cite{2009} and Spivak {\it et al.}\cite{2002}.

The advantage of (\ref{G2})-(\ref{new44}) is two fold. Firstly, the poor asymptotic behaviours (\ref{new11}) and (\ref{new33}) for the original integrands  are  improved respectively  as
\bbe\label{nn1}
\frac{[2\nu \!+\!(k\!+\!\nu)e^{-2kh}][\e^{k(z\!+\!\zeta)}\!+\!\e^{ k(z-\zeta-2h)}\!+\!\e^{k(-z\!+\!\zeta-2h)}\!+\!\e^{ -k(z\!+\!\zeta\!+\!4h)}]J_0(kR)}{(1\!+\!e^{-2kh})(k\tanh kh-\nu-\i\mu ) }
\sim \frac{2\nu\sqrt{2}\cos (kR-\frac{\pi}4)}{k
\sqrt{\pi k R}}
\bee
and
\begin{align}
&\frac{(2\nu^2\!+\!(k\!+\!2\nu)(k\!+\!\nu)e^{-2kh}) (\e^{k(z+\zeta)}\!+\!\e^{ k(z-\zeta-2h)}\!+\!\e^{k(\zeta-z-2h)}\!+\!\e^{ -k(z+\zeta+4h)})  J_1(kR)}{(1\!+\!e^{-2kh})(k\tanh kh\!-\!\nu\!-\!\i\mu ) }
\nonumber\\
&\sim \frac{2\nu^2\sqrt{2}\cos (kR-\frac{3\pi}4)}{
k\sqrt{\pi k R}}\label{nn2}
\end{align}
on the free surface $z=0$ and $\zeta=0$, so that the wave integrations in (\ref{G2}) and (\ref{new44}) are convergent for $k\to \infty$.
Secondary, the  formalization  implies  that  (\ref{G2}) and (\ref{new44}) with the limit $\mu\to 0$ can be replaced by those with  a value $0<\mu <<1$ so that  the wave integrals are  on the straight line $0<k<\infty$ rather than on the curve shown in Figure \ref{f1}.

Thus  evaluation of the Green function can be obtained by directly integrating  a smooth function involving a small value of $\mu>0$. The numerical result is accurate in comparison with that given by John's  series (\ref{GG3}).  The direct integration evaluation is efficient in application  to wave body motions from very good agreement  of the present method results with  the semi-analytic results of  Wang \cite{W1986}, Hulme\cite{Hu1982} and Linton \cite{Linton1999}.

As given in \cite{WehausenLaitone}, the original formulas (\ref{G1}) or  (\ref{new22}) on the curved integration line can be written as a Cauchy-principle-value (CPV) integral plus $\pi \i$ times the residue of the integrand at the pole $k=k_0$. A CPV integral part close to the pole $k_0$  may  be obtained by the method of Monacella \cite{1967} using the property
\bbe
(CPV)\int^{k_0+a}_{k_0-a} \frac{dk}{k-k_0} =0,
\bee
although the  function $\frac1{k\tanh(kh)-\nu}$ is even with respect to $k$. However, the integration close to the infinity is not convergent on the free surface due to (\ref{new11}) and (\ref{new33}). Therefore,  (\ref{G1}) or  (\ref{new22}) cannot be integrated directly  in the  rigorous  analysis manner. Thus it is necessary to use other approximations (see, for example, \cite{N1985,Jon}).

\section{New formulation of the free surface Green function $G$}

It is convenient to use the
inverse  Hankel transformation or the inverse Fourier transformation in the polar  coordinate system  of  the horizontal plane (see, for example, \cite[p. 384]{Wa})
\bbe
\frac1{\sqrt{R^2+z^2}}
&=&  \int^\infty_0\frac1k\e^{kz}J_0(kR) kdk= {\mathcal H}^{-1}(\frac1k\e^{kz}),\,\,\, z<0.\label{G3}
\bee

It follows from (\ref{G3}) that, for $z+\zeta<0$,
\be
\frac1{r_1}+\frac1{r_2}+\frac1{r_3}+\frac1{r_4}&=& \int^\infty_0\left(\e^{k(z+\zeta)}+\e^{ k(z-\zeta-2h)}+\e^{k(-z+\zeta-2h)}+\e^{ -k(z+\zeta+4h)}\right)J_0(kR)dk
\\
&=&\int^\infty_04\e^{-2kh}\cosh k(\zeta+h)\cosh k(z+h)J_0(kR)dk.
\ee
This yields that
\be
 \lefteqn{\int_L\frac{2(\nu +k)\e^{-kh}\cosh k(\zeta+h)\cosh k(z+h)}{k\sinh kh-\nu \cosh kh }J_0(kR)dk-(\frac1{r_1}+\frac1{r_2}+\frac1{r_3}+\frac1{r_4})}
\\ &=&  \int_L\left(\frac{(\nu +k)}{(1+e^{-2kh})(k\tanh kh-\nu ) }-1\right)4\e^{-2kh}\cosh k(\zeta+h)\cosh k(z+h)J_0(kR)dk
\\ &=&  \int_L\frac{[2\nu +(k+\nu)e^{-2kh}][\e^{k(z+\zeta)}+\e^{ k(z-\zeta-2h)}+\e^{k(-z+\zeta-2h)}+\e^{ -k(z+\zeta+4h)}]}{(1+e^{-2kh})(k\tanh kh-\nu ) }J_0(kR)dk.
\ee
Note that exponential functions and the Bessel function are continuous along the positive real line and thus independent of the integral pass change in Figure \ref{f1}.  Hence it remains  to check the analytical behaviour of the
   following singular integral on the lower half circle around $k=k_0$:
   \def\Im{\rm Im}
\bbe
\int_{ |k-k_0|=\epsilon, \,\Im (k)\leq 0} \frac{dk}{k\tanh kh-\nu}
=\frac1{\tanh k_0h}\int_{ |k-k_0|=\epsilon, \,\Im (k)\leq 0} \frac{dk}{k-k_0}=\frac{\i \pi}{\tanh k_0h}\label{G44}
\bee
for a given constant $\epsilon>0$ sufficiently small.
In contrast, we have   the integral
\bbe  \int^{k_0+\epsilon}_{k_0-\epsilon} \frac{dk}{k\tanh kh-\nu-\i\mu}&\approx&\frac1{\tanh k_0h}\int^{k_0+\epsilon}_{k_0-\epsilon} \frac{dk}{k-k_0-\frac{\i\mu}{\tanh k_0h}}\nonumber
\\
&=&\frac1{\tanh k_0h}\ln \frac{\epsilon\tanh k_0h-\i \mu}{-\epsilon\tanh k_0h-\i\mu}\nonumber
 \\
 &=&
\frac{\i \pi}{\tanh k_0h} +\frac1{\tanh k_0h}\ln \left(1+\frac{-2\i \mu}{\epsilon\tanh k_0h+\i\mu}\right).\label{newG}
\bee
This together with (\ref{G44}) implies that, for any small $\epsilon>0$,
\be \int_{ |k-k_0|=\epsilon, \,\Im(k)\leq 0} \frac{dk}{k\tanh kh-\nu}=\lim_{\mu\to 0+}\int^{k_0+\epsilon}_{k_0-\epsilon} \frac{dk}{k\tanh kh-\nu-\i\mu}
\ee
and thus we have the desired formulation (\ref{G2}).

\begin{figure}[h]
\centering 
\includegraphics[width=.4\textwidth,height=.4\textwidth]{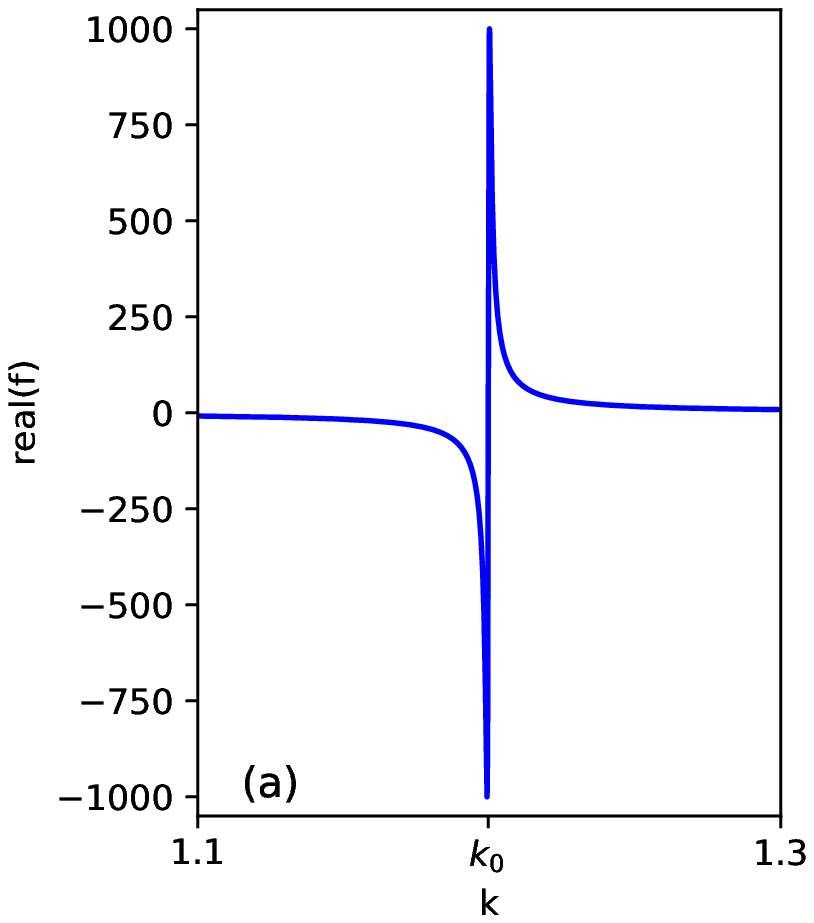}
\includegraphics[width=.4\textwidth,height=.4\textwidth]{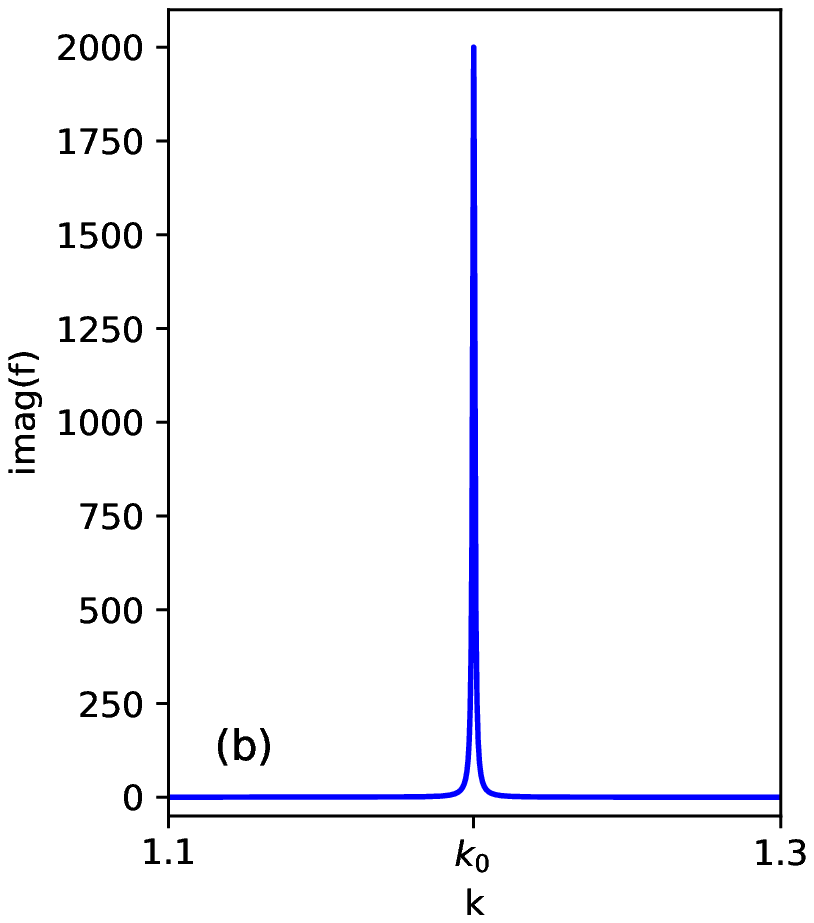}
\includegraphics[width=.4\textwidth,height=.4\textwidth]{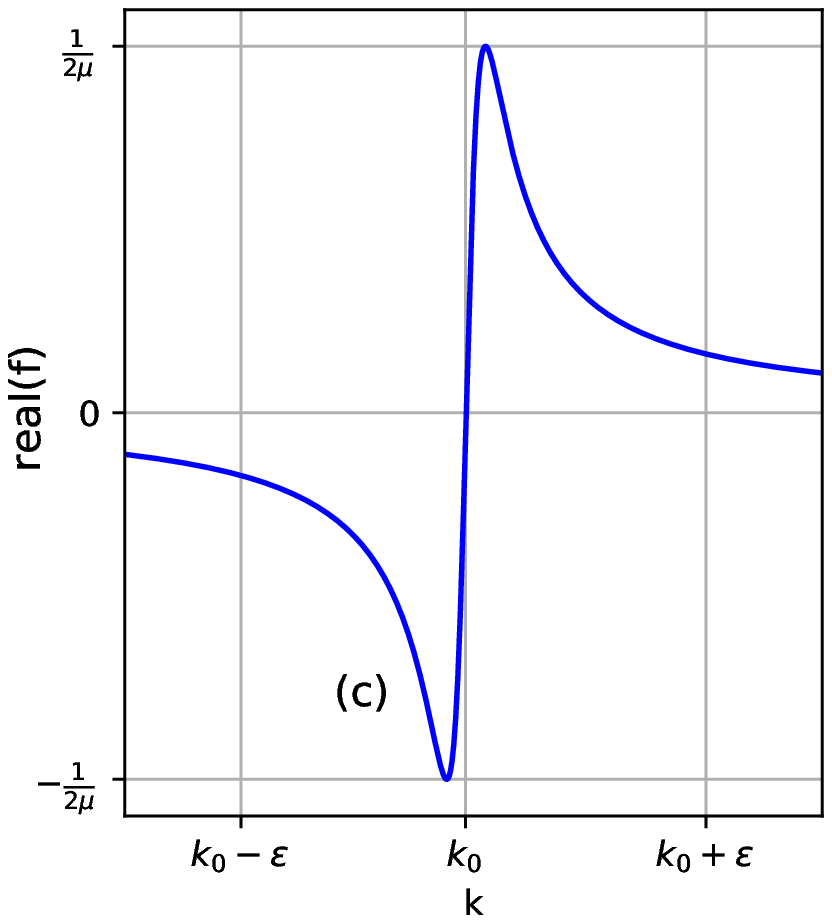}
\includegraphics[width=.4\textwidth,height=.4\textwidth]{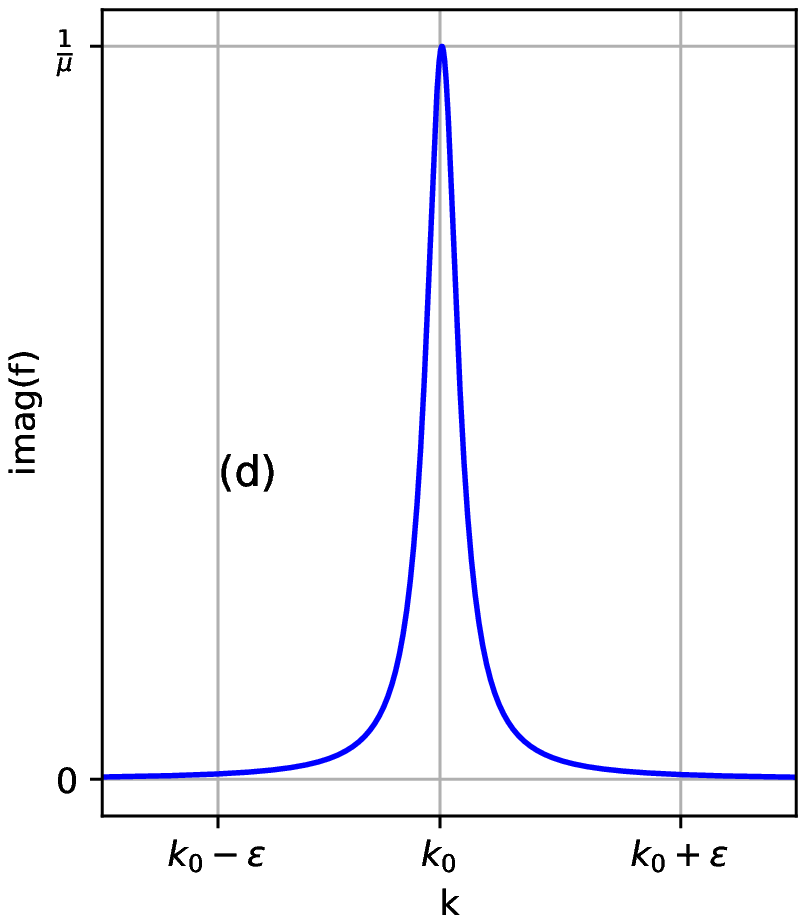}

\caption{(a) and (b): The function $f(k)=\frac1{k\tanh kh-\nu-\i\mu}$ for $h=1$, $\nu=1$, $k_0=1.19965$ and $\mu=0.0005$;  (c) and (d): profile of the smooth function $f$ around $k=k_0$;}
\label{f2}
\end{figure}
The behaviour of the function  $\frac1{k\tanh kh-\nu-\i\mu}$ is displayed in Figure \ref{f2}. Although this function is becoming sharper and unbounded as $\mu\to 0$, its real part close to $k_0$ is symmetric with respect to the  point $k=k_0$, while the imaginary part tends to the dirac delta function at the vicinity of the root $k=k_0$. Thus the integral of the function around $k_0$ or the area bounded by the function and the real line as given by (\ref{newG})  is always meaningful.
Therefore, the integral of the real part in the vicinity of $k_0$ is zero and the integral of the imaginary part around $k_0$  remains  constant for $\mu$ sufficiently small.
However, as shown in Figure \ref{f2}, the sharpness of the function increases  when the parameter  $\mu>0$ decreases. Thus denser meshgrid points around $k_0$ are necessary for smaller $\mu>0$.

Moreover, the dispersion relation (\ref{dis}) becomes the deep water dispersion relation $k-\nu=0$ when $h\nu > 7$, since $\tanh(k_0h) =1$ numerically for $k_0h>7$. Thus compared with the infinite water depth case,
 the main difference of the finite water depth case  is defined  by the integral on the integral domain $[0, 7/\nu]$.

Next, we  consider the vertical derivative of $K$. By (\ref{G3}), it shows that
\begin{align*}
\lefteqn{\p_\zeta K-2\nu(\frac1{r_1}-\frac1{r_2}+\frac1{r_3}-\frac1{r_4})}\\
&= \!\!\!\int_L\left(\frac{[2\nu +(k+\nu)e^{-2kh}]k}{(1\!+\!e^{-2kh})(k\tanh kh\!-\!\nu) }\!-\!2\nu\right)(\e^{k(z+\zeta)}\!-\!\e^{ k(z-\zeta-2h)}\!+\!\e^{k(\zeta-z-2h)}\!-\!\e^{ -k(z+\zeta+4h)})J_0(kR)dk
\\&=  \int_L\frac{2\nu^2 +(k+2\nu)(k+\nu)e^{-2kh}}{(1+e^{-2kh})(k\tanh kh-\nu ) }(\e^{k(z+\zeta)}\!-\!\e^{ k(z-\zeta-2h)}\!+\!\e^{k(\zeta-z-2h)}\!-\!\e^{ -k(z+\zeta+4h)})J_0(kR)dk.
\end{align*}
Hence, by (\ref{G2}), we have
\begin{align}
\lefteqn{\p_\zeta G- \p_\zeta(\frac1r+\frac1{r_0}+\frac1{r_1}+\frac1{r_2}+\frac1{r_3}+\frac1{r_4})-2\nu(\frac1{r_1}-\frac1{r_2}+\frac1{r_3}-\frac1{r_4})}\label{aG2}
\\&= \nonumber\lim_{\mu\to 0+} \int^\infty_0\frac{2\nu^2 +(k+2\nu)(k+\nu)e^{-2kh}}{(1+e^{-2kh})(k\tanh kh-\nu-\i\mu ) }(\e^{k(z+\zeta)}\!-\!\e^{ k(z-\zeta-2h)}\!+\!\e^{k(\zeta-z-2h)}\!-\!\e^{ -k(z+\zeta+4h)})J_0(kR)dk.
\end{align}

Finally, for the Bessel function $J_1(s) =-\frac{d J_0(s)}{ds}$, we have
\begin{align} \nonumber\lefteqn{\p_R K+ \int^\infty_02\nu (\e^{k(z+\zeta)}\!+\!\e^{ k(z-\zeta-2h)}\!+\!\e^{k(\zeta-z-2h)}\!+\!\e^{ -k(z+\zeta+4h)}) J_1(kR)dk }\\
&=- \!\!\int_L\!\!\left(\frac{[2\nu \!+\!(k\!+\!\nu)e^{-2kh}]k}{(1\!+\!e^{-2kh})(k\tanh kh\!-\!\nu ) }-2\nu\right)(\e^{k(z+\zeta)}\!+\!\e^{ k(z-\zeta-2h)}\!+\!\e^{k(\zeta-z-2h)}\!+\!\e^{ -k(z+\zeta+4h)}) J_1(kR)dk \nonumber
\\ &= -\!\!\int_L\!\!\frac{2\nu^2\!+\!(k\!+\!2\nu)(k\!+\!\nu)e^{-2kh} }{(1\!+\!e^{-2kh})(k\tanh kh\!-\!\nu ) }(\e^{k(z+\zeta)}\!+\!\e^{ k(z-\zeta-2h)}\!+\!\e^{k(\zeta-z-2h)}\!+\!\e^{ -k(z+\zeta+4h)})  J_1(kR)dk.\label{NNN}
\end{align}
With the use of integration by parts and (\ref{G3}), we have
\be
\lefteqn{\int^\infty_02\nu (\e^{k(z+\zeta)}\!+\!\e^{ k(z-\zeta-2h)}\!+\!\e^{k(\zeta-z-2h)}\!+\!\e^{ -k(z+\zeta+4h)}) J_1(kR)dk }
\\&=&-\frac{2\nu}R\int^\infty_0  (\e^{k(z+\zeta)}\!+\!\e^{ k(z-\zeta-2h)}\!+\!\e^{k(\zeta-z-2h)}\!+\!\e^{ -k(z+\zeta+4h)})  dJ_0(kR)
\\ &=& \frac{8\nu}R+ \frac{2\nu}R\int^\infty_0 \left( (z+\zeta)\e^{k(z+\zeta)}+(z-\zeta-2h)\e^{ k(z-\zeta-2h)} \right.
\\&&\left.+(\zeta-z-2h)\e^{k(\zeta-z-2h)}-(z+\zeta+4h)\e^{ -k(z+\zeta+4h)}\right)  J_0(kR) d k
\\ &=&\frac{8\nu}R+ \frac{2\nu}R\left(\frac{z+\zeta}{r_1}+\frac{z-\zeta-2h}{r_2}+\frac{\zeta-z-2h}{r_3}+\frac{-z-\zeta-4h}{r_4}\right)
\\ &=&\frac{2\nu R}{r_1(r_1+|z+\zeta|)}+\frac{2\nu R}{r_2(r_2+|z-\zeta-2h|)}+\frac{2\nu R}{r_3(r_3+|\zeta-z-2h|)}+\frac{2\nu R}{r_4(r_4+|z+\zeta+4h|)}.
\ee
This together with (\ref{G44}) and (\ref{NNN}) implies  that
\begin{align*} \lefteqn{\p_R K+ \frac{2\nu R}{r_1(r_1\!+\!|z\!+\!\zeta|)}\!+\!\frac{2\nu R}{r_2(r_2\!+\!|z\!-\!\zeta\!-\!2h|)}\!+\!\frac{2\nu R}{r_3(r_3\!+\!|\zeta\!-\!z\!-\!2h|)}\!+\!\frac{2\nu R}{r_4(r_4\!+\!|z\!+\!\zeta\!+\!4h|)} }\\
\\ &= -\!\!\lim_{\mu\to 0}\int^\infty_0\!\!\frac{2\nu^2\!+\!(k\!+\!2\nu)(k\!+\!\nu)e^{-2kh} }{(1\!+\!e^{-2kh})(k\tanh kh\!-\!\nu\!-\!\i\mu ) }(\e^{k(z+\zeta)}\!+\!\e^{ k(z-\zeta-2h)}\!+\!\e^{k(\zeta-z-2h)}\!+\!\e^{ -k(z+\zeta+4h)})  J_1(kR)dk.
\end{align*}
This gives the formulation for the horizontal derivative of the Green function
\begin{align} \lefteqn{\p_R G - \partial_R \left(\frac1{r}+\frac1{r_0}+\frac1{r_1}+\frac1{r_2}+\frac1{r_3}+\frac1{r_4}\right)}\nonumber\\
 &= \nonumber -\!\!\lim_{\mu\to 0+}\int^\infty_0\!\!\frac{2\nu^2\!+\!(k\!+\!2\nu)(k\!+\!\nu)e^{-2kh} }{(1\!+\!e^{-2kh})(k\tanh kh\!-\!\nu\!-\!\i\mu ) }(\e^{k(z+\zeta)}\!+\!\e^{ k(z-\zeta-2h)}\!+\!\e^{k(\zeta-z-2h)}\!+\!\e^{ -k(z+\zeta+4h)})  J_1(kR)dk
\\ \label{aG3}
&-\left( \frac{2\nu R}{r_1(r_1\!+\!|z\!+\!\zeta|)}\!+\!\frac{2\nu R}{r_2(r_2\!+\!|z\!-\!\zeta\!-\!2h|)}\!+\!\frac{2\nu R}{r_3(r_3\!+\!|\zeta\!-\!z\!-\!2h|)}\!+\!\frac{2\nu R}{r_4(r_4\!+\!|z\!+\!\zeta\!+\!4h|)}\right).
\end{align}

When $h\to \infty$, the formulation  reduces to the infinite  depth Green function
\begin{align}
&G=\frac1r+\frac1{r_1}+\lim_{\mu\to0+} \int^\infty_0\frac{2\nu \e^{k(z\!+\!\zeta)}J_0(kR)dk}{k-\nu-\i\mu  } , \label{aaG1}
\\
&\p_\zeta G= \p_\zeta(\frac1r+\frac1{r_1})+\frac{2\nu}{r_1}\label{aaG2}
+\lim_{\mu\to 0+} \int^\infty_0\frac{2\nu^2\e^{k(z+\zeta)}J_0(kR)}{k-\nu-\i\mu  }dk,
\\
& \p_R G = \partial_R \left(\frac1{r}+\frac1{r_1}\right) - \frac{2\nu R}{r_1(r_1\!+\!|z\!+\!\zeta|)}-\!\!\lim_{\mu\to 0+}\int^\infty_0\!\!\frac{2\nu^2\e^{k(z+\zeta)}  J_1(kR)}{k\!-\!\nu\!-\!\i\mu  }dk.
 \label{aaG3}
\end{align}

\section{Direct  integration of the Green function}
\def\km{k_{\rm max}}
Let $K^\mu$ denote the wave integral of (\ref{G2}) involving $\mu>0$. By  (\ref{G2}), (\ref{aG2}) and (\ref{aG3}), we have
$$K=\lim_{\mu \to 0+} K^\mu, \,\,\,\,\partial_\zeta K=\lim_{\mu \to 0+} \partial_\zeta K^\mu, \,\,\,\,\,\partial_R K=\lim_{\mu \to 0+} \partial_R K^\mu.$$
Therefore we may numerically take
$$K\approx K^\mu,\,\,\, \partial_\zeta K\approx \partial_\zeta K^\mu,\,\,\, \partial_R K\approx \partial _R K^\mu$$ for $\mu>0$ sufficiently small.

Note that the wave integral     $K^\mu$ and those of $\p_\zeta K^\mu$ and $\p_R K^\mu$  are convergent  even for the limit case $z=\zeta=0$ or the integration $\int^\infty_{k_{\rm max}}$ of the corresponding integrands is uniformly small for a large $k_{\rm max}$.  The infinite integration domain $0<k<\infty$ is thus truncated by finite integration domain $0<k<k_{\rm max}$ for a suitable number $k_{\rm max}$.
   For simplicity, we may use the direct integration in the approximation manner
\begin{align} K& \approx \sum^{N_k}_{j=0}\int^{k_{j\!+\!1}}_{k}\frac{[2\nu \!+\!(k\!+\!\nu)e^{-2k_jh}][\e^{k(z\!+\!\zeta)}\!+\!\e^{ k(z-\zeta-2h)}\!+\!\e^{k(-z\!+\!\zeta-2h)}\!+\!\e^{ -k(z\!+\!\zeta\!+\!4h)}]}{(1\!+\!e^{-2kh})(k\tanh kh-\nu-\i\mu ) J_0(kR)}dk\nonumber
\\
&\approx  \sum^{N_k}_{j=1}\frac{[2\nu \!+\!(k_j\!+\!\nu)e^{-2k_jh}][\e^{k_j(z\!+\!\zeta)}\!-\!\e^{ k_j(z-\zeta-2h)}\!+\!\e^{k_j(-z\!+\!\zeta-2h)}\!-\!\e^{ -k_j(z\!+\!\zeta\!+\!4h)}]J_0(k_jR)}{(1\!+\!e^{-2k_jh})} \nonumber
\\
&\cdot \frac1{\tanh k_{j+1}h}\ln \frac{k_{j+1}\tanh k_{j+1}h-\nu-\i\mu } {k_{j}\tanh k_{j+1}h-\nu -\i\mu}.\label{aaaG1}
\end{align}
Here  $\{k_j\}^{N_k}_{j=1}$ represents  a set of  meshgrid points of the truncation  domain $0<k<k_{\rm max}$ and is sufficiently dense so that the numerator of integrand together with the exponential function  $e^{-2kh}$ and the hyperbolic function $\tanh (kh)$ are  approximately   constant on   $[k_j, k_{j+1}]$.

Similarly, we have the following evaluation
\begin{align}
\lefteqn{\p_\zeta K -2\nu(\frac1{r_1}-\frac1{r_2}+\frac1{r_3}-\frac1{r_4})}\nonumber
\\&\approx  \nonumber \sum_{j=1}^\infty\!\int^{k_{j+1}}_{k_j}\!\!\!\!\!\frac{2\nu^2 \!+\!(k\!+\!2\nu)(k\!+\!\nu)e^{-2kh}}{(1\!+\!e^{-2kh})(k\tanh kh\!-\!\nu\!-\!\i\mu ) }(\e^{k(z+\zeta)}\!-\!\e^{ k(z-\zeta-2h)}\!+\!\e^{k(\zeta-z-2h)}\!-\!\e^{ -k(z+\zeta+4h)})J_0(kR)dk
\\
&\approx  \sum^{N_k}_{j=1}\frac{[2\nu^2 \!+\!(k_j\!+\!2\nu)(k_j\!+\!\nu)e^{-2k_jh}][\e^{k_j(z\!+\!\zeta)}\!+\!\e^{ k_j(z-\zeta-2h)}\!+\!\e^{k_j(-z\!+\!\zeta-2h)}\!+\!\e^{ -k_j(z\!+\!\zeta\!+\!4h)}]}{(1\!+\!e^{-2k_jh})} \nonumber
\\
&\cdot \frac1{\tanh k_{j+1}h}\ln \frac{k_{j+1}\tanh k_{j+1}h-\nu-\i\mu } {k_{j}\tanh k_{j+1}h-\nu -\i\mu}J_0(k_jR)\label{aaaG2}
\end{align}
and
\begin{align} \lefteqn{\p_R K  +\left( \frac{2\nu R}{r_1(r_1\!+\!|z\!+\!\zeta|)}\!+\!\frac{2\nu R}{r_2(r_2\!+\!|z\!-\!\zeta\!-\!2h|)}\!+\!\frac{2\nu R}{r_3(r_3\!+\!|\zeta\!-\!z\!-\!2h|)}\!+\!\frac{2\nu R}{r_4(r_4\!+\!|z\!+\!\zeta\!+\!4h|)}\right)}\nonumber
\\
&\approx  -\sum^{N_k}_{j=1}\frac{[2\nu^2 \!+\!(k_j\!+\!2\nu)(k_j\!+\!\nu)e^{-2k_jh}][\e^{k_j(z\!+\!\zeta)}\!+\!\e^{ k_j(z-\zeta-2h)}\!+\!\e^{k_j(-z\!+\!\zeta-2h)}\!+\!\e^{ -k_j(z\!+\!\zeta\!+\!4h)}]}{(1\!+\!e^{-2k_jh})} \nonumber
\\
&\cdot \frac1{\tanh k_{j+1}h}\ln \frac{k_{j+1}\tanh k_{j+1}h-\nu-\i\mu } {k_{j}\tanh k_{j+1}h-\nu -\i\mu}J_1(k_jR).\label{aaaG3}
\end{align}

To improve the accuracy of the  evaluation formulas (\ref{aaaG1})-(\ref{aaaG3}), we may use $\tanh(k_0h)$ instead of  $\tanh(k_{j+1}h)$  whenever $k_0\in [k_{j},k_{j+1}]$.

With the use of (\ref{aaaG3}),  we have the  evaluation for the horizontal partial derivatives
\bbe
\p_\xi K= \p_RK \frac{\xi-x}R  \,\,\,\mbox{ and }\,\,\, \p_\eta K= \p_RK \frac{\eta-y}R. \label{nnn1}
\bee

Here we do not use integral approximation methods such as trapezoidal rule and Simpson rules, which are not designed for dealing with the integration of a sharp function close to the dirac delta function. In order to cancel the huge positive integration area with the huge  negative integration area displayed in Figure 2(a,c) and to compute the constant integration area bounded by the dirac delta like function in Figure 2(b,d) in a tiny integral interval
$[k_0-\epsilon, k_0+\epsilon]$, it is convenient to integrate  the sharp function  $\frac{1}{ k\tanh(k_0h)-\nu -\i \mu}$ around the pole $k=k_0$, under the condition that  the meshgrid is sufficiently dense.

\begin{figure}[h]
\centering 
\includegraphics[width=.48\textwidth,height=.35\textwidth]{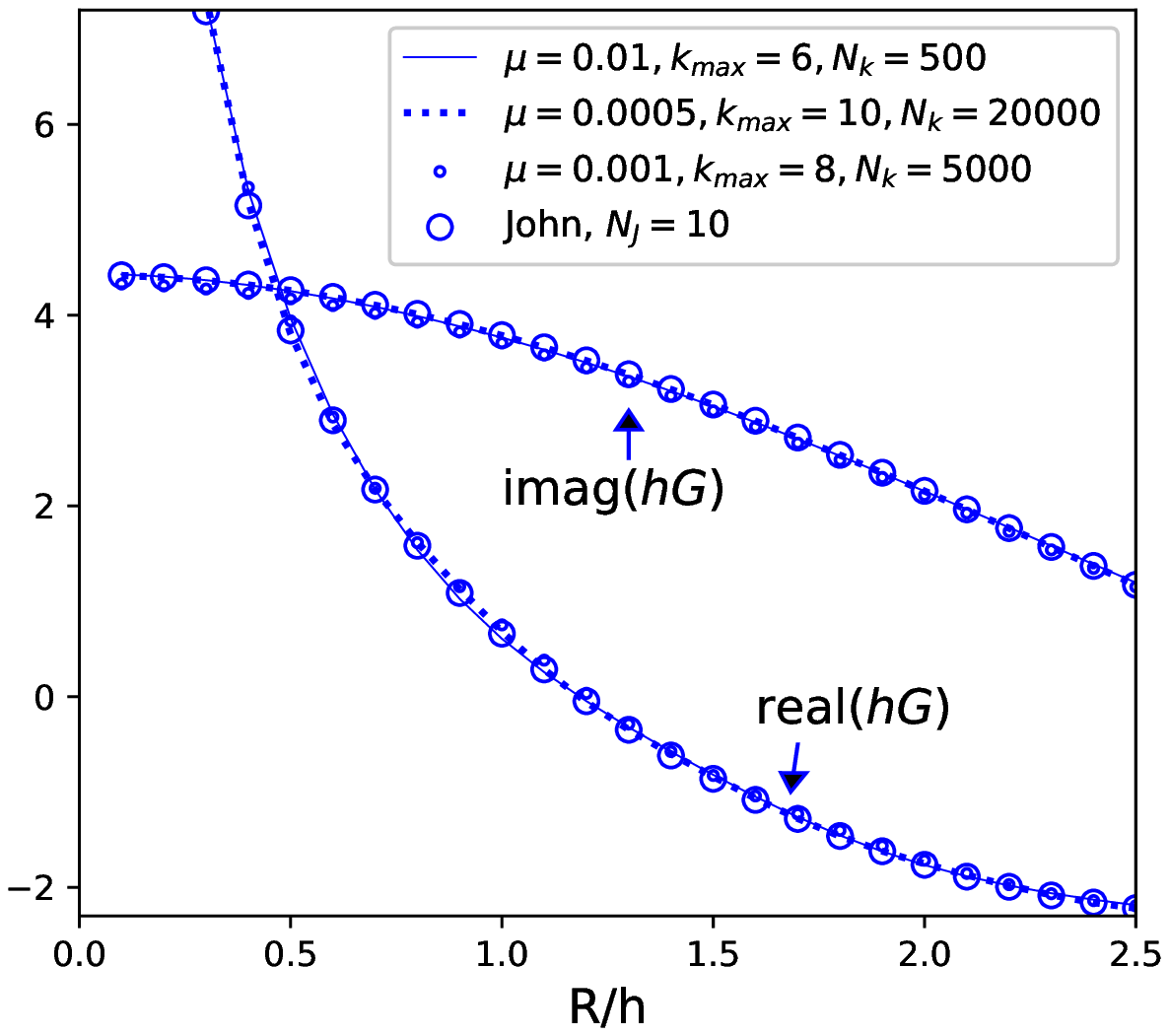}
\includegraphics[width=.48\textwidth,height=.35\textwidth]{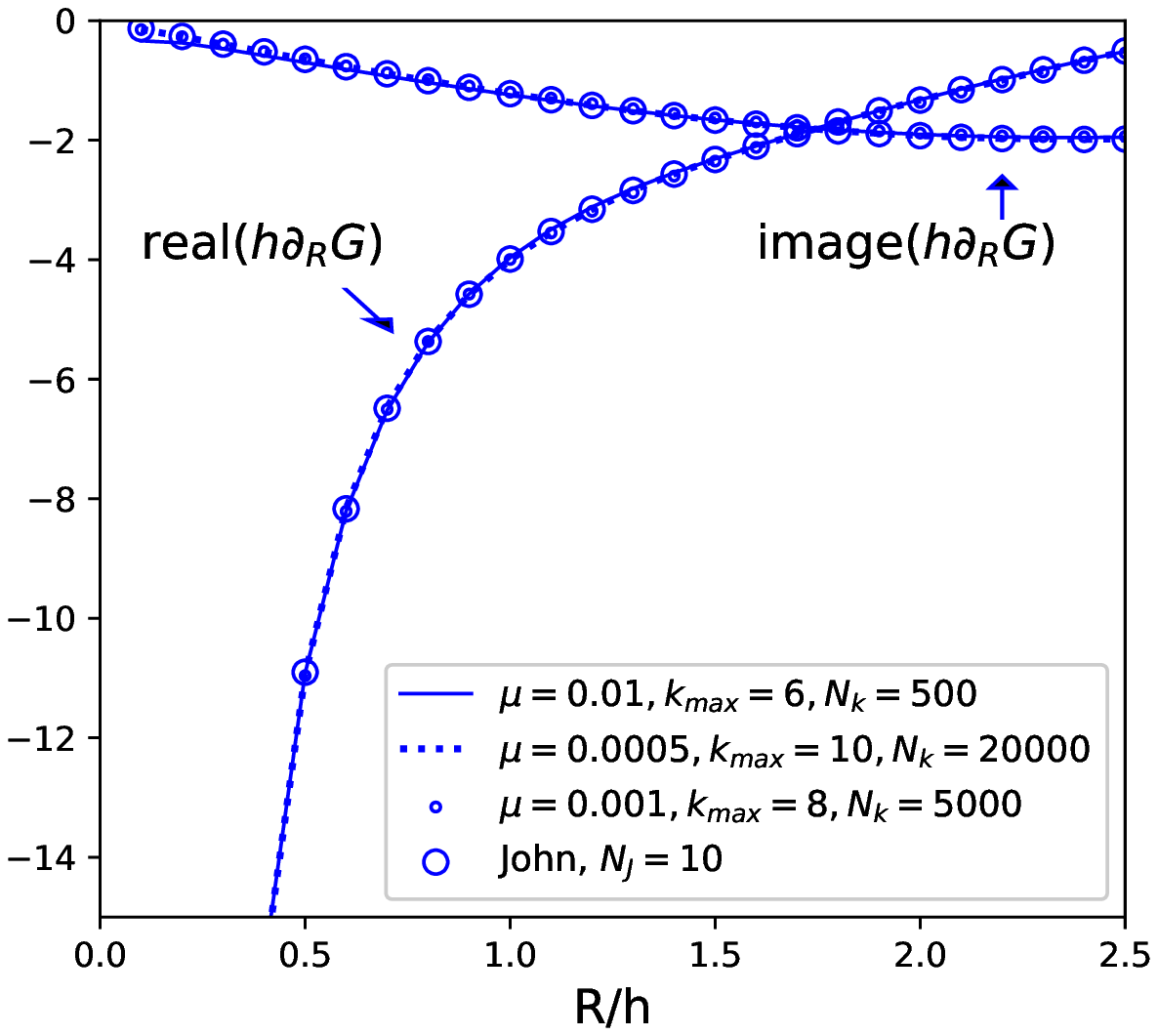}\vspace{2mm}

\caption{Comparison of present integration method and the John's expansion method (\ref{GG3}) on  evaluating  the Green function $G$ and its horizontal derivative $\partial_R G$  at the condition  $h\nu=0.5$. }
\label{f3}
\end{figure}

In the computation for Figure \ref{f3},  we use the uniform meshgrid
\bbe k_j = \frac{(j-1)k_{\rm max}}{N_k}
\bee
for simplicity,
where the integer  $N_k$ increases with  $\mu$. To derive the suitable results in Figure \ref{f3}, we have to take  $N_k=20000$ for $\mu=0.0005$ and $N_k=5000$ for $\mu=0.001$. However, it is also possible to obtain suitable result for $\mu=0.01$  and $N_k=500$. For the result of John's expansion, we take $N_J=10$ as the expansion result remains almost  unchanged for $N_J>10$.
  The roots $k_0$ and $\i k_n$ of  the dispersion equation are  determined  by the Newton iteration method. The Bessel functions $J_0$, $Y_0$ and $K_0$ are evaluated by the polynomial  expansions in \cite{Abr}. The non-dimensional horizontal variable  $R/h$ on the interval $[0, 2.5]$  takes 25 ordinates.  In computing $G$ and  $\p_RG$ by using John's expansion formula (\ref{GG33})  via a Fortran 90 code in an PC of i5-4460 CPU@3.20GHz, the elapsed time  is less than 1 second. On the other hand, the elapsed time for  the computation of $G$ and $\p_RG$ through the formulas (\ref{myG2}), (\ref{aaaG1}) and (\ref{aaaG3}) with  $\mu=0.001$ and $N_k=5000$  is less than 1 second as well.

 Actually, $K^\mu$ remains little changed  for small $\mu \leq 0.01$. For displaying purpose, Figure \ref{f3} shows that  $G$ and $\p_\zeta G$  with $\mu=0.001$ and $0.0005$ in the present evaluation  are almost the same with those given by John's series (\ref{GG3}). Thus we mainly use the value $\mu=0.001$ in our computations.

\begin{figure}[h!]
\centering
\includegraphics[width=.48\textwidth,height=.33\textwidth]{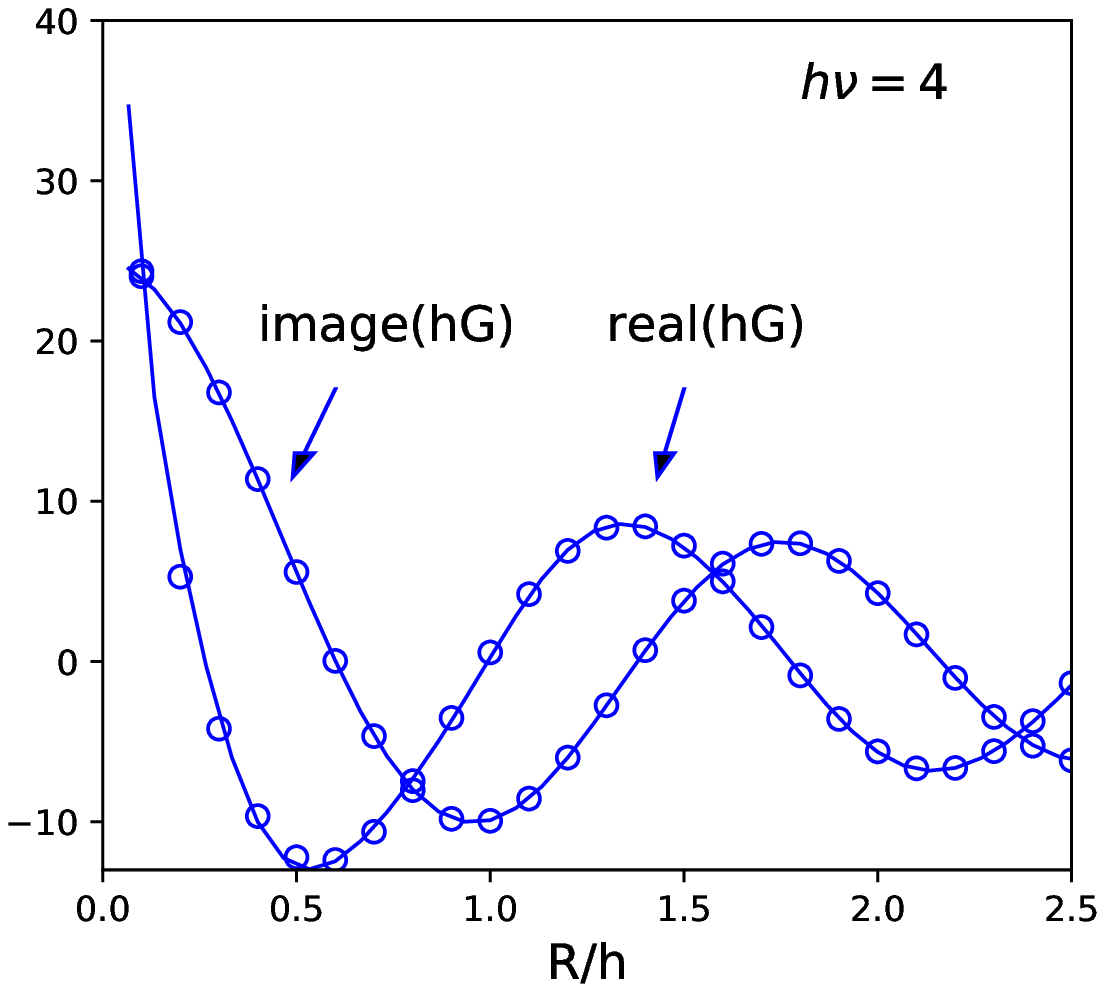}
\includegraphics[width=.48\textwidth,height=.33\textwidth]{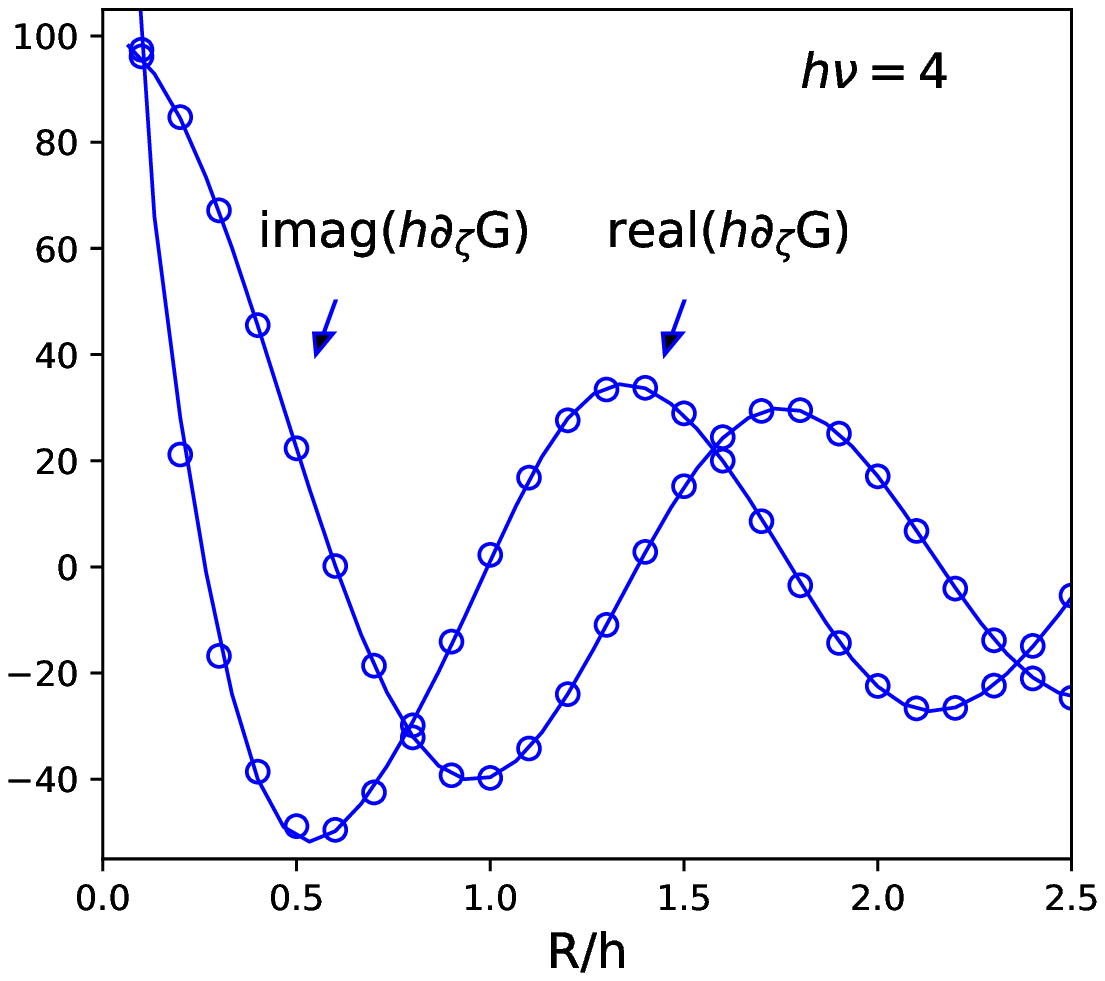}
\vspace{-2mm}

\includegraphics[width=.48\textwidth,height=.33\textwidth]{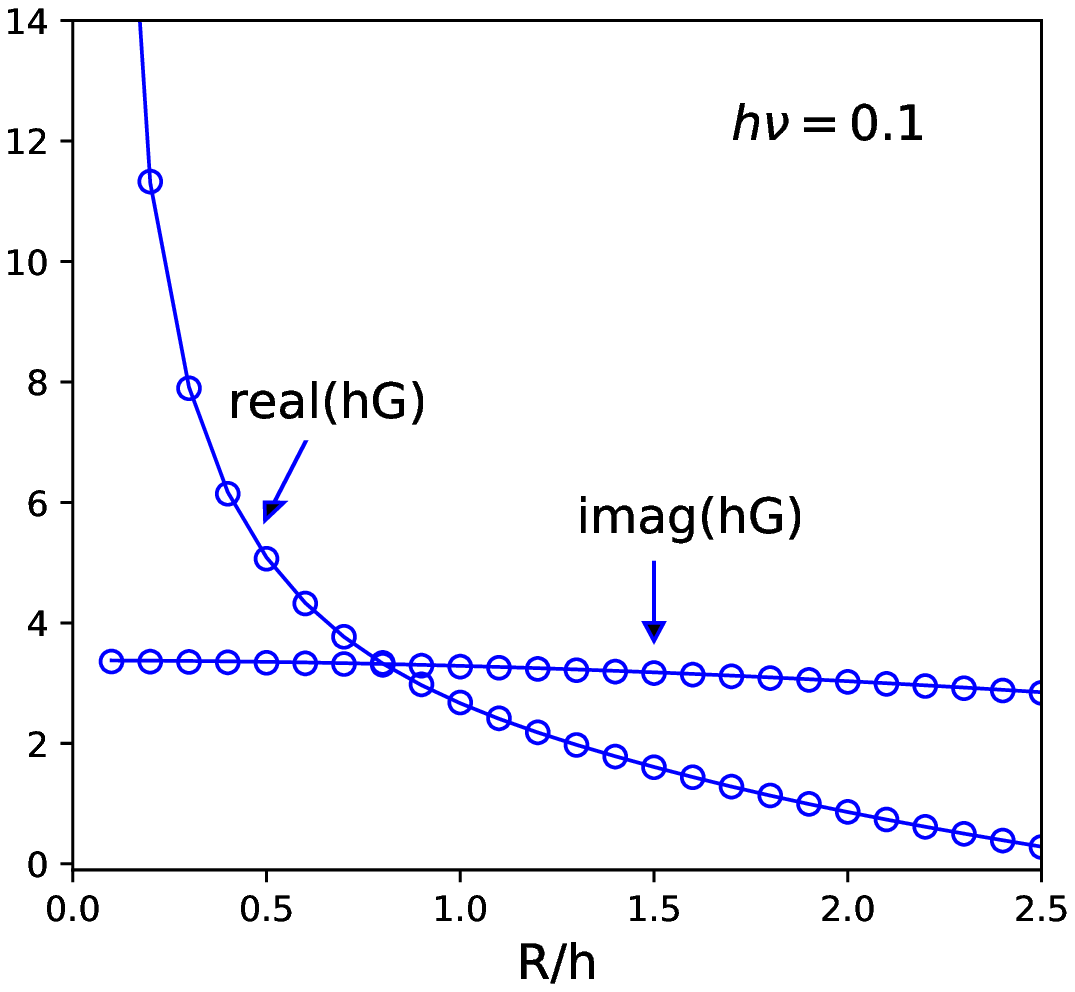}
\includegraphics[width=.48\textwidth,height=.33\textwidth]{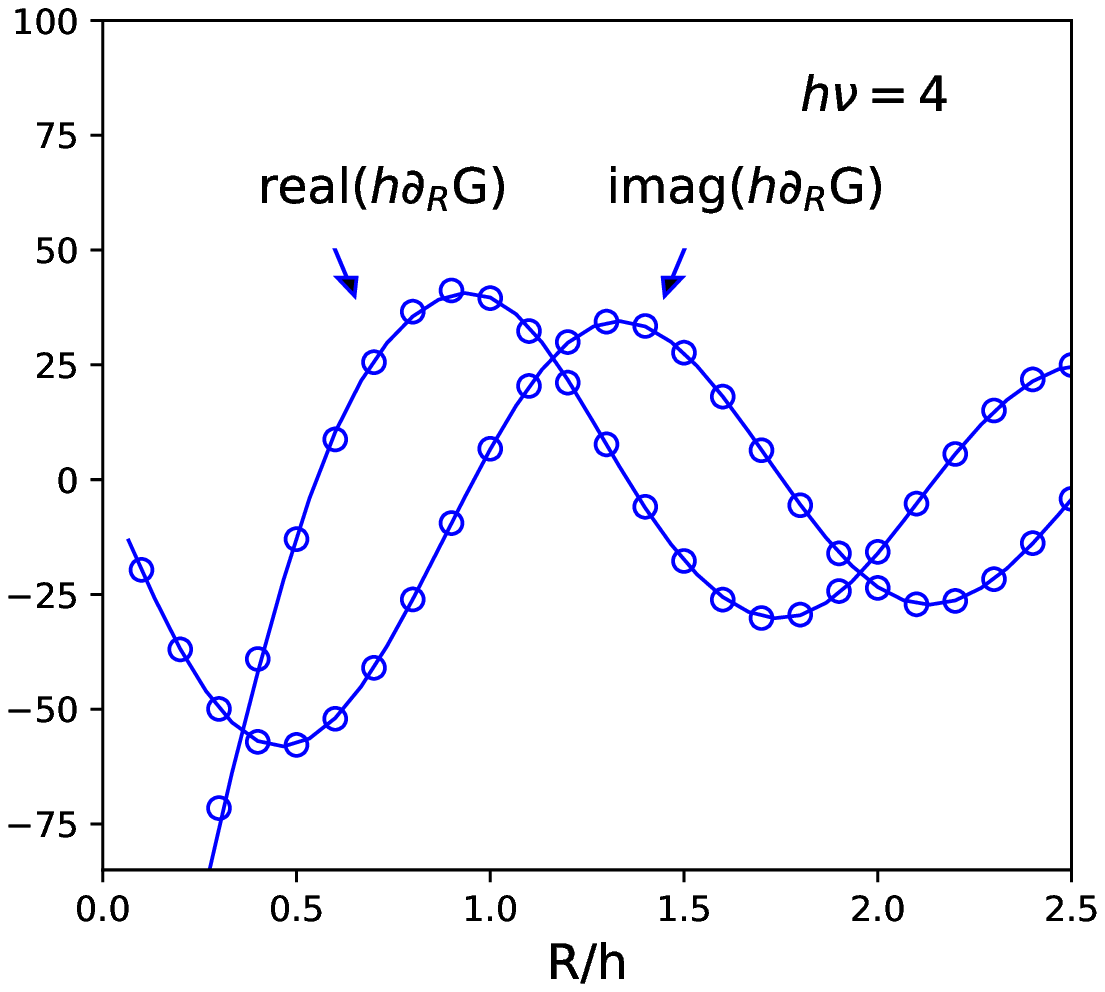}
\vspace{-2mm}

\includegraphics[width=.48\textwidth,height=.33\textwidth]{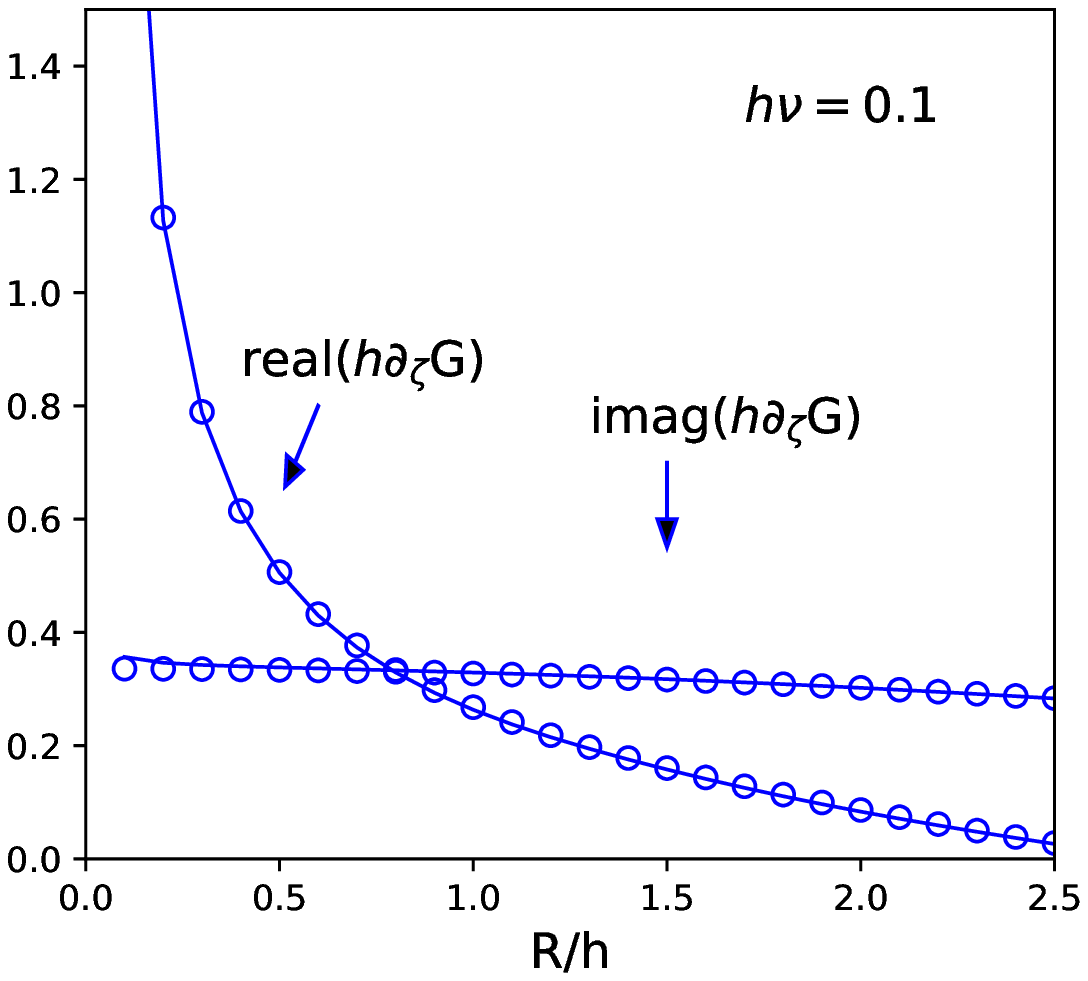}
\includegraphics[width=.48\textwidth,height=.33\textwidth]{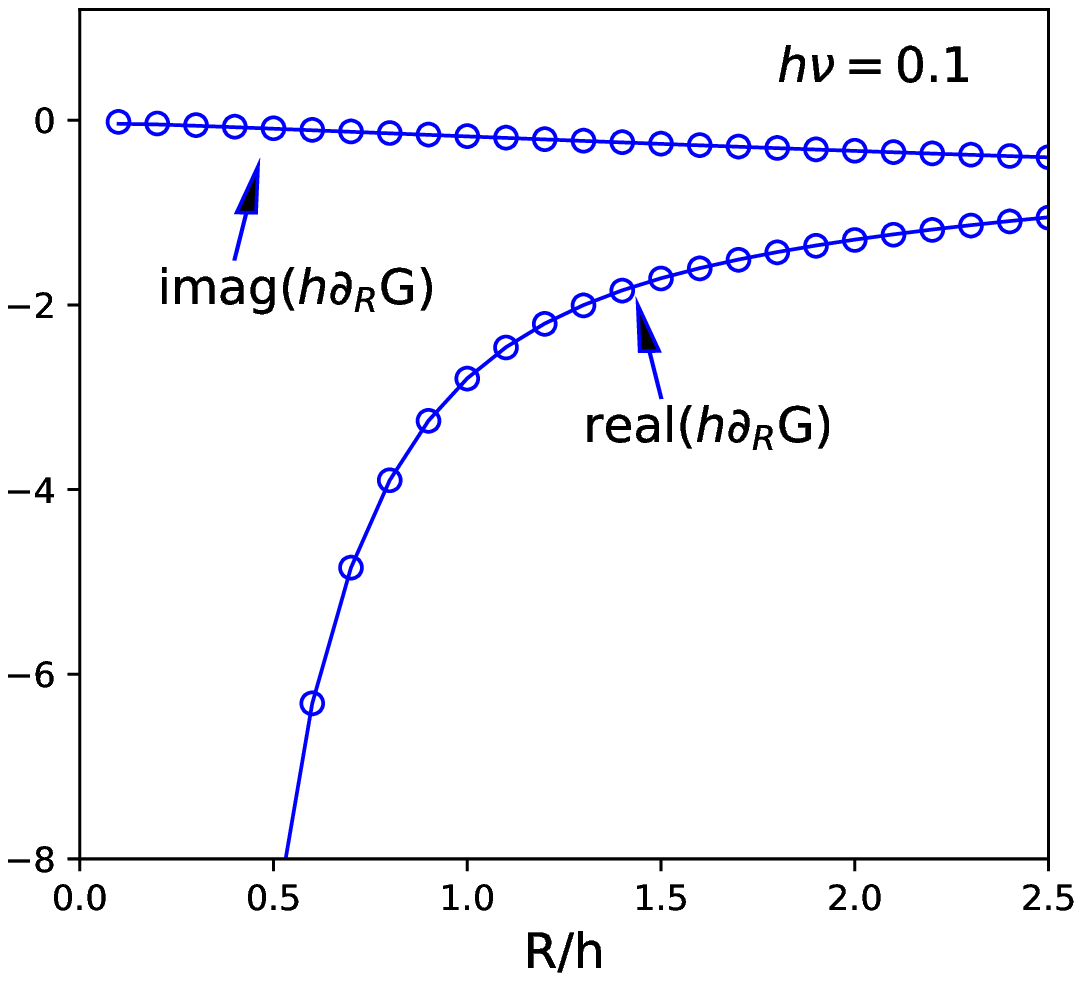}
\caption{Comparison  of the numerical Green function results  given respectively  by the present method (solid lines, $\mu=0.001$) and   John's method \cite{Jon} (circle lines, $N_J=10$) defined by (\ref{GG33}). Moreover, for the present method,   $k_{\rm max}=8$ and $N_k=5000$ whenever $h\nu=4$, but $k_{\rm max}=6$ and $N_k=3000$ whenever $h\nu=0.1$.}
\label{f4}
\end{figure}

With the use of the approximation (\ref{aaaG1})-(\ref{aaaG3}), the evaluation of the Green function becomes simple but robust.
To help understanding the present Green function evaluation, selected numerical results are displayed in Figure \ref{f4}, which shows  the accuracy of the present evaluation  in comparison with John's expansion (\ref{GG3}) for the non-dimensional wave number $h\nu$ at the moderate value $h\nu=4$ and the small value $h\nu=0.1$.

\section{The Green function method in the wave body motion problem}

Consider a three-dimensional body    undergoing periodic oscillatory motion with a constant frequency $\omega$  in the fluid  $-h < z <0$, the velocity potential of the linearized oscillatory fluid motion problem can be represented as
\bbe \Phi = \R(\phi \e^{-\i \omega t}) \label{new3}
\bee
where  the   complex potential  $\phi$ is a harmonic function satisfying the free surface boundary condition (\ref{con2}) and the water bed boundary condition (\ref{con3}). Thus $\phi$ is determined by the boundary integral equation
\bbe \label{bb}\phi(\q) +\frac1{4\pi}\int_{S}\phi(\pp)\, \n_{\pp} \!\cdot\! \nabla_p G(\q,\pp)
 \d S_{\pp}&=&\frac1{4\pi}\int_{S}G(\q,\pp)\,\n_{\pp}\!\cdot\!  \nabla \phi(\pp) \d S_{\pp}
\bee
for the field point $\q$ in the fluid domain together with  the linear body boundary condition
\bbe \n_{\pp}\!\cdot\! \nabla \phi(\pp)= -\i\omega  n_\alpha\,\,\mbox{ on } S.\label{rad}
\bee
Here $S$ is the average wetted body  surface  and $\n_{\pp}=\n(\pp)=(n_1,n_2,n_3)(\pp)$ represents the normal vector field at $\pp \in S$ and pointing into the fluid domain.
 The body undergoes heave motion for $\alpha =3$, sway motion for $\alpha =2$ and surge motion for $\alpha =1$.

When  the field point in the fluid domain $\D$  tends to the wetted body boundary $S$,  eq. (\ref{bb}) reduces to  the boundary integral equation
\bbe \phi(\q) +\frac1{4\pi}  \lim_{\q' \in \D, \q'\to \q}  \int_{S}  \phi(\pp)\,\n_{\pp}\!\cdot\! \nabla_{\pp}G(\q',\pp) \d  S_{\pp} = \frac{1}{4\pi} \int_{S} G(\q,\pp) \,\n_{\pp}\!\cdot\! \nabla \phi(\pp) \d S_{\pp},\,\,\, \q\in S,\label{boundarya}\bee
which is approximated by  the finite boundary element discretisation equation  system
\begin{align} \phi(\q_{i,j})& +\frac1{4\pi} \sum_{I=1}^N\sum_{J=1}^M \phi(\pp_{I,J})\lim_{\q' \in \D, \q'\to \q_{i,j}}  \int_{\panel_{I,J}}  \n_{\pp}\!\cdot\! \nabla_{\pp} G(\q',\pp) \d  S_{\pp}\nonumber
 \\
 &= \frac{1}{4\pi} \sum_{I=1}^N\sum_{J=1}^M(- \i )\omega n_\alpha (\q_{I,J})\int_{\panel_{I,J}} G(\q_{i,j},\pp)  \d S_{\pp}\label{boundary},\,\,\, 1\leq i\leq N,\,\,\,\, 1\leq j\leq M
\end{align}
by using the surface mesh discretisation
$$ S\approx \sum_{I=1}^N\sum_{J=1}^M \panel_{I,J}$$
defined by mesh grid points $\pp_{I,J}$ with  $I=1,...,N+1$ and $J=1,..., M+1$. Here $\q_{I,J}$ presents the centre point of $\panel_{I,J}$.


Let $|\panel_{I,J}|$  denote the area of $\panel_{I,J}$.
The influence coefficients can be calculated as
\begin{align}
\lefteqn{\int_{\panel_{I,J}} G(\q_{i,j},\pp)  \d S_{\pp}}\nonumber\\
&\approx\frac{1}{4\pi}\int_{\panel_{I,J}}\left(\frac1{|\q_{i,j}-\pp|}+\sum_{l=0}^4 \frac1{r_l(\q_{i,j},\pp)}\right)\d S_{\pp}
+
\frac{1}{4\pi}K(\q_{i,j},\pp_{I,J}) |\panel_{I,J}|
\label{new1}
\end{align}
and
\begin{align}
 \lefteqn{\lim_{\q' \in \D, \q\to \q_{i,j}}  \int_{\panel_{I,J}}  \n_{\pp}\!\cdot\! \nabla_{\pp}G(\q,\pp) \d  S_{\pp}- \lim_{\q\in\D, \q\to \q_{i,j}}\int_{\panel_{I,J}}\n_{\pp}\!\cdot\!\nabla_\pp  \frac1{|\q- \pp|} \d S_{\pp}}\nonumber
 \\ &\approx
  \int_{\panel_{I,J}}\sum_{l=0}^4\n_{\pp}\!\cdot\! \nabla_\pp  \frac1{r_l(\q_{i,j},\pp)} d S_{\pp}+\!\!\n_{\pp}\!\cdot\! \nabla_\pp  K(\q_{i,j}, \pp_{I,J}) |\panel_{I,J}|.
\label{new2}
\end{align}
Here $\nabla_\pp K$ is evaluated by (\ref{aaaG2})-(\ref{nnn1}),
\begin{align}
\lim_{\q\in\D, \q\to \q_{i,j}}\int_{\panel_{I,J}}\n_\pp\!\cdot\! \nabla_\pp \frac1{|\q- \pp|}\d S_{\pp}=\int_{\panel_{I,J}}\n_\pp\!\cdot\!\nabla_\pp\frac1{|\q_{i,j}- \pp|}\d S_{\pp}, \,\,\,\,\mbox{ when } (i,j)\neq (I,J),
\label{new3a}
\end{align}
and
\begin{align}
\lim_{\q\in\D, \q\to \q_{I,J}}\int_{\panel_{I,J}}\n_\pp\!\cdot \!\nabla_\pp \frac1{|\q- \pp|} \d S_{\pp}
=\lim_{z\to 0+}\int_{S_{I,J}}\left.\frac{\partial}{\partial \hat \zeta} \frac{1 }{\sqrt{{\hat \xi}^2+{\hat\eta}^2+(z-\hat \zeta)^2}}\right|_{\hat \zeta=0} \d \hat \xi\d \hat \eta=2\pi,
\label{new4}
\end{align}
where we have used the coordinate transformation by transform $\panel_{I,J}$ in the global coordinate system on to  the panel $S_{I,J}$ centered at $(0,0,0)$ in a local coordinate system $(\hat \xi,\hat \eta,\hat \zeta)$.

The panel integrals of $\frac1{|\q_{i,j}-\pp|}$,  $\frac1{r_l}$ and  their normal derivatives  are  given  by the Hess-Smith quadrilateral integral method
 \cite{HS,HS2,Newman1986}. Therefore, with the use of   the Green function approximation  (\ref{aaaG1})-(\ref{aaaG3}) in (\ref{new1}) and (\ref{new2}), the evaluation of the influence coefficients is obtained. Thus the algebraic equation system (\ref{boundary}) can be numerically solved
by  the Gaussian elimination scheme.

\section{Added mass and damping coefficients}
\begin{figure}[h!]
\centering 
\centering
\includegraphics[width=.48\textwidth,height=.48\textwidth]{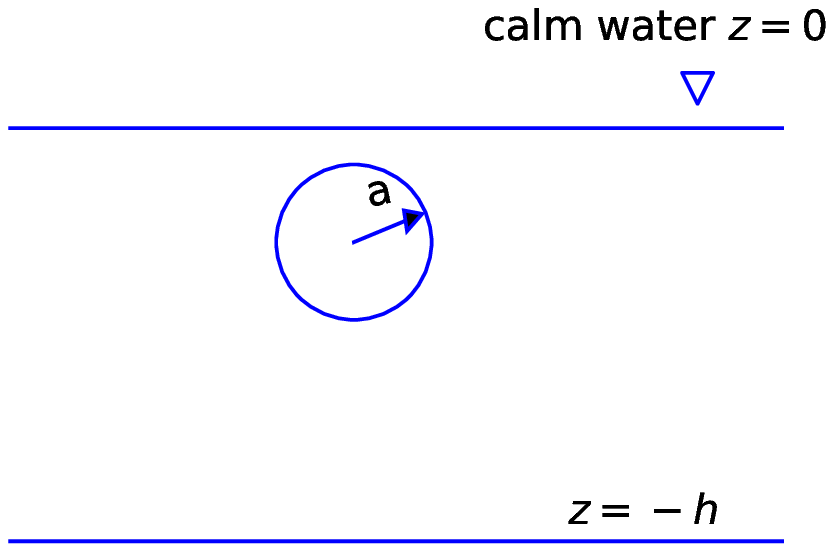}
\includegraphics[width=.48\textwidth,height=.48\textwidth]{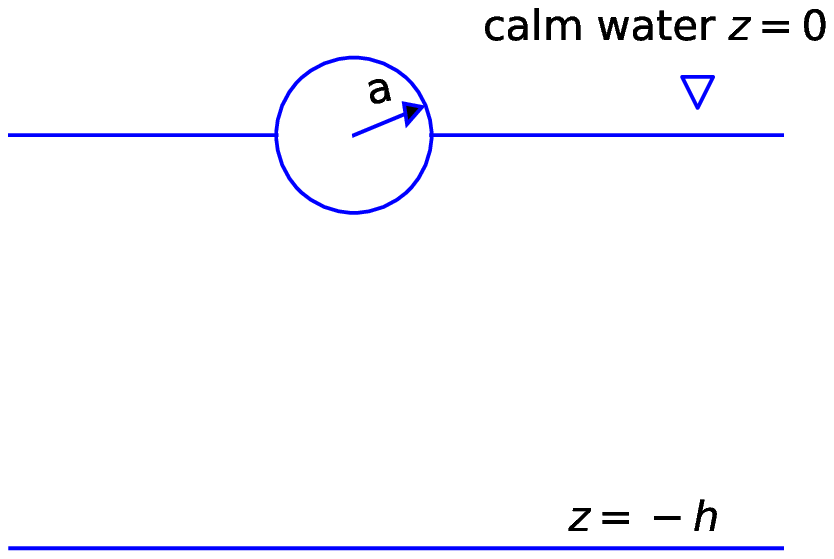}
\vspace{-17mm}

\caption{Cross sections ($y=0$) of a submerged sphere of radius $a$ centred at $(0,0,-1.5a)$ and a floating sphere of radius $a$. }
\label{f5}
\end{figure}

To validate  the Green function evaluation in practice and understand wave induced loading to an oscillating  body in waves, we calculate numerically added mass and damping coefficients for the Green function method in wave body motions.
Consider firstly a sphere of radius $a$ submerged in the fluid $-h<z<0$. This  sphere   is centred at $(0,0,-h_0)$ with $h_0/a=1.5$ (see Figure \ref{f5})

For the numerical velocity potential solution $\phi=\phi_\alpha$ ($\alpha=1, 2, 3$) of the boundary value problem (\ref{rad}) and (\ref{boundarya}),  the linear hydrodynamic  pressure  is expressed as
\be
p_\alpha=-\rho \frac{\p \Phi_\alpha}{\p t} = \omega\rho\R\left( \i \phi_\alpha \e^{-\i \omega t}\right)
\ee
for $\rho$ the fluid density.
This  defines the hydrodynamic wave force exerted on the average wetted body surface $S$:
$$F_{\alpha,\alpha}= \int_S p_\alpha n_\alpha dS$$
and the non-dimensional added mass and damping coefficients $A_{\alpha,\alpha}$ and $B_{\alpha,\alpha}$:
\bbe  A_{\alpha,\alpha}+\i B_{\alpha,\alpha} &=& \frac 1{\omega V } \int_{S}\i\phi_\alpha n_\alpha \d S\label{add}
\approx \frac 1{\omega V } \sum_{i=1}^N \sum_{j=1}^M \i\phi_\alpha(\q_{i,j}) n_{\alpha}(\q_{i,j})|\panel_{i,j}|.\bee
Here  $V$ is the volume of the moving body with the wetted body surface $S$.
Especially,   $V= \frac43\pi a^3$  for the submerged sphere and   $V= \frac23\pi a^3$  for the floating hemisphere.

\begin{figure}[h!]
\centering 
\includegraphics[width=.99\textwidth,height=.28\textwidth]{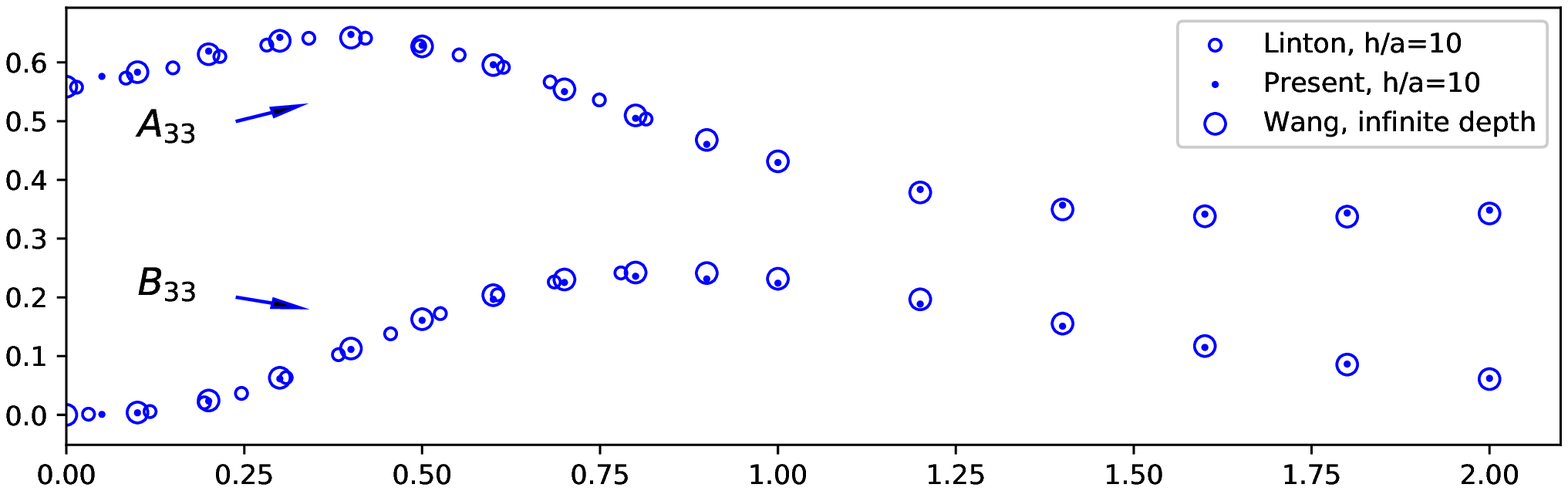}
\includegraphics[width=.99\textwidth,height=.28\textwidth]{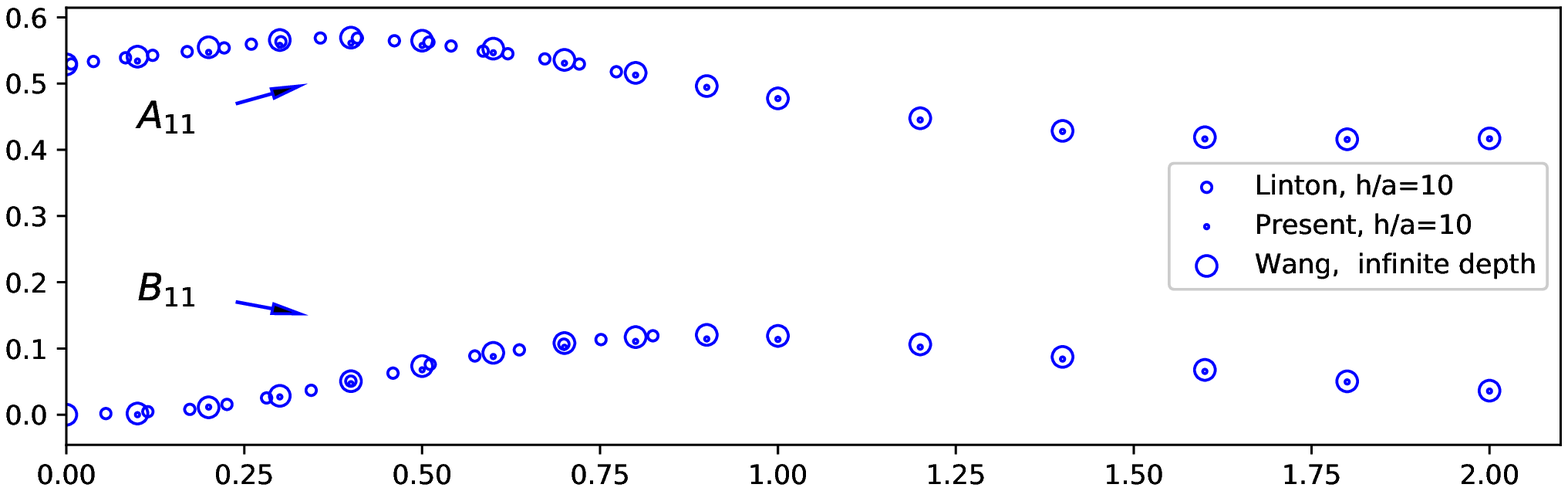}
\includegraphics[width=.99\textwidth,height=.28\textwidth]{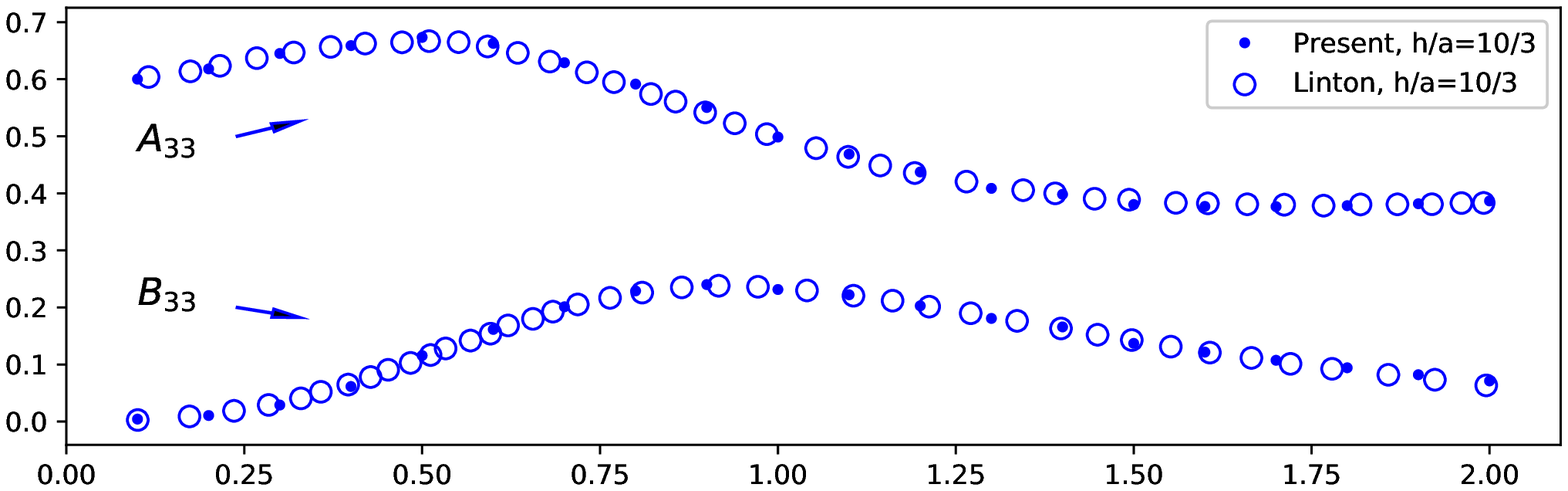}
\includegraphics[width=.99\textwidth,height=.28\textwidth]{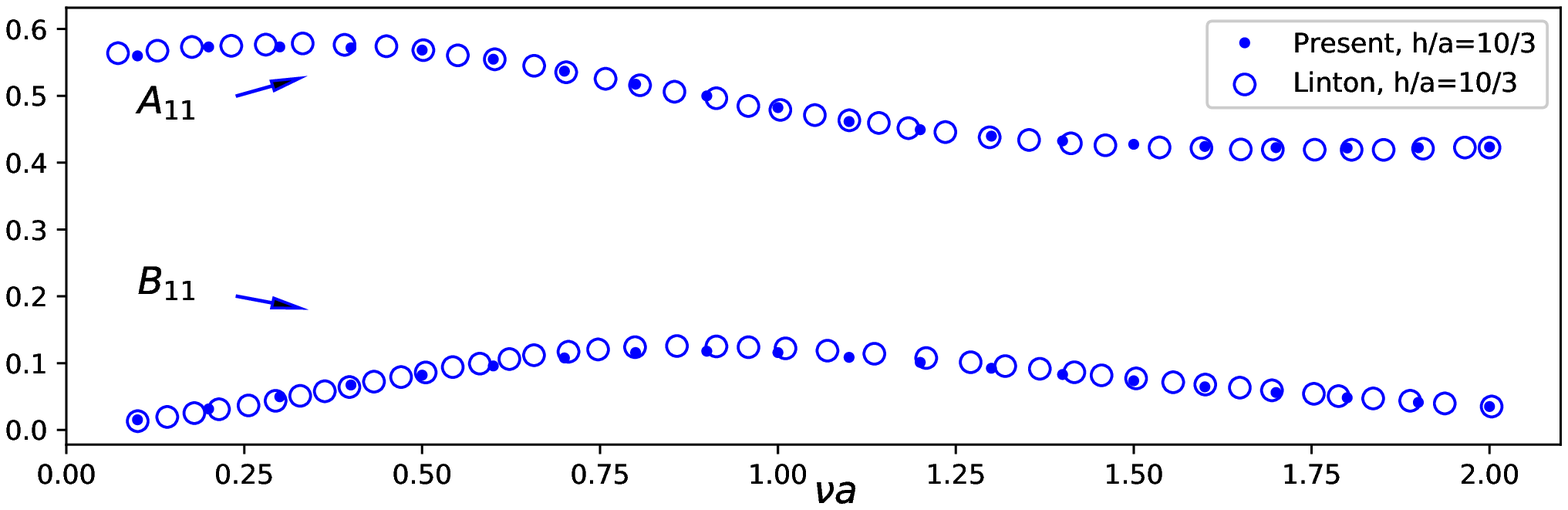}
\caption{Added mass and damping coefficients of heaving and surging  sphere of radius $a$ submerged at the depth of $h_0=1.5a$ and derived by the present method and the semi-analytical methods of Wang \cite{W1986} for  the infinite water depth $-h\to -\infty$  and Linton \cite{Linton1991} for   finite water bed $z=-h$ values, which are only available in \cite{Linton1991} for $\nu a< 0.8$ when $h/a=10$. }
\label{f6}
\end{figure}

Selected results of added mass and damping coefficients of  heave and surge motions at different $h/a$ values  are displayed in Figure \ref{f6}.
For the deep water case  $h/a=10$, the present method results coincide with the semi-analytical results of Linton \cite{Linton1991} , where  sway rather than surge motion is calculated. However, for the radial symmetric body, the sway motion is the same with the surge motion. Actually, for the submerged sphere oscillating at the water depth $h_0/a=1.5$, the influence of the water bed $h/a=10$ is negligible due to the comparison of
the results  with the semi-analytical results of Wang \cite{W1986} in infinite water depth situation. Figure \ref{f6} also shows good agreement for the present method and  Linton \cite{Linton1991} results at $h/a=10/3$.

\begin{figure}[h!]
\centering 
\includegraphics[width=.99\textwidth,height=.28\textwidth]{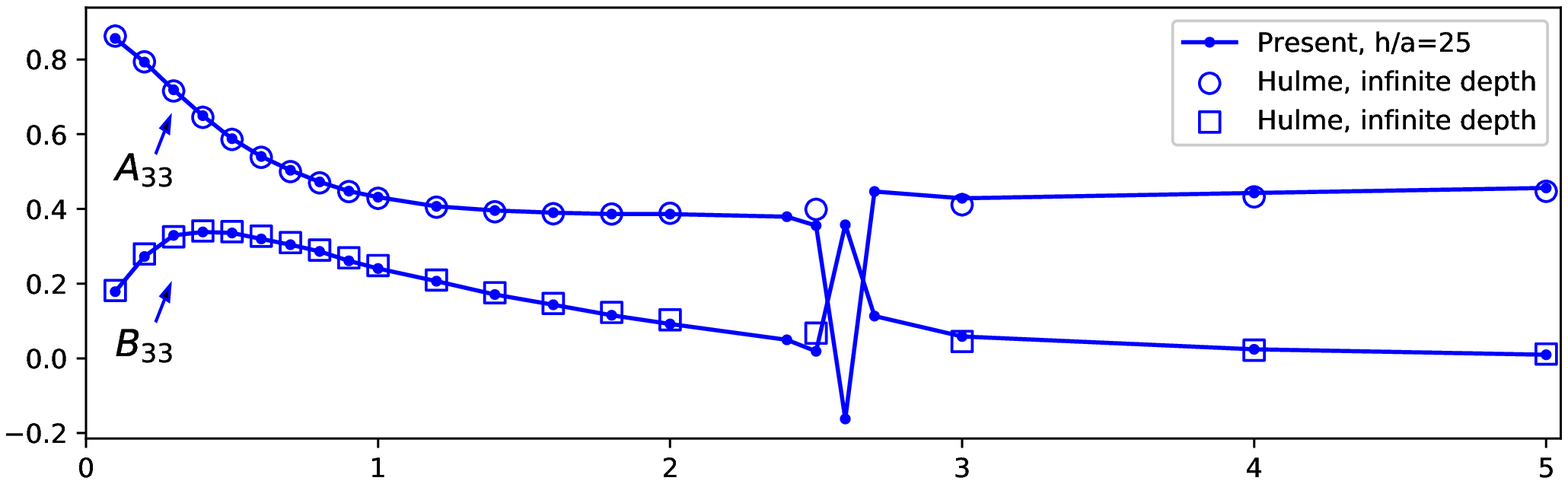}
\includegraphics[width=.99\textwidth,height=.28\textwidth]{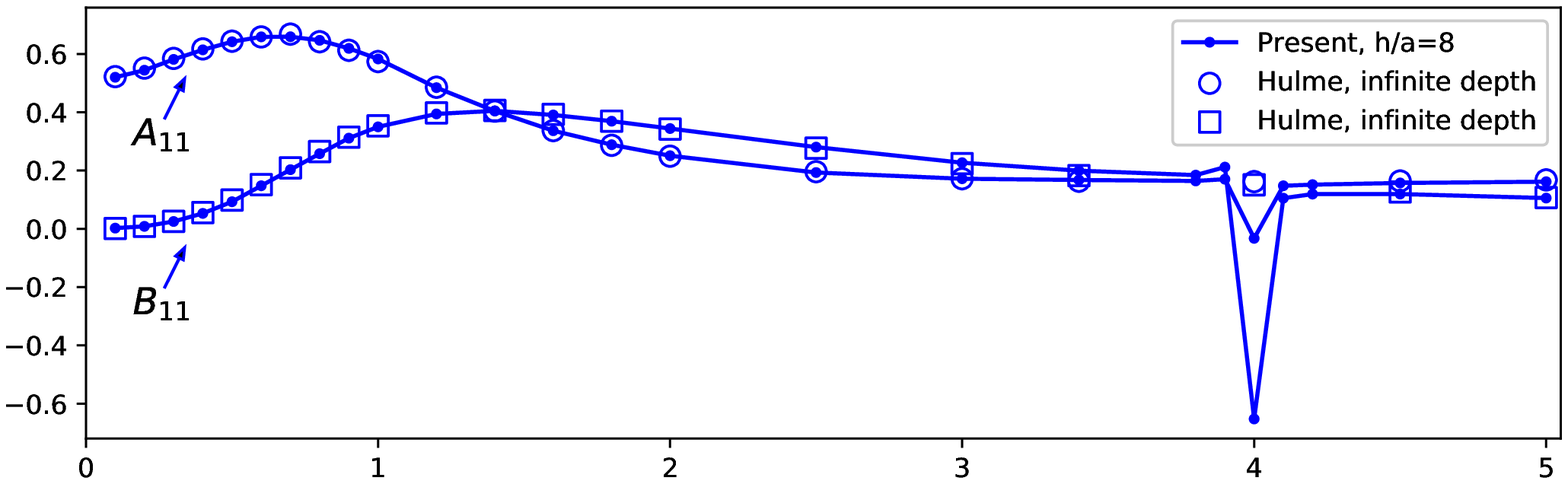}

\caption{Added mass and damping coefficients of heaving and surging  floating hemisphere of radius $a$  derived by the present method and the semi-analytical methods of Hulme  \cite{Hu1982} for  the infinite water depth $-h\to -\infty$.}
\label{f7}
\end{figure}

Next, for the comparison with the semi-analytic results of Hulme \cite{Hu1982} at infinite water depth, we consider a floating  hemisphere (see Figure \ref{f5})  undergoing respectively heave and surge motions in deep water depth. The comparison is illustrated in Figure \ref{f7}. The present method results exhibit irregular frequencies at $\nu a$ at the vicinity of $\nu a =2.6$ for heaving hemisphere motion and of $\nu a = 4$ for the surging hemisphere motion.  As is well known (see,  Frank \cite{Frank1967} John \cite{Jon})  that the combination of panel method and free surface Green function gives rise to irregular
frequencies in a high frequency range when a floating body undergoes oscillatory motions. Various methods exist (see, for
example, Lee and Sclavounos \cite{Lee1989}  Ursell \cite{U1981}, Lee {\it et al.} \cite{Lee1996} and Zhu and Lee \cite{ZhuLee1994} ) to remove non-physical irregular frequencies. We may also use  simple  correction method by  interpolating  regular frequency data as given by the author \cite{Chen2015}  so that the smooth interpolation data for the deep water depths present excellent agreement with the results of Hulme \cite{Hu1982}. It should be noted that the surging hemisphere oscillates  in the horizontal direction and thus is not very sensitive with the water depth. Here we take $h/a=8$ in Figure \ref{f7}. Actually, the numerical results remain the same for $h/a=4$. This case is comparable with the Smith effect.  However,  the heaving hemisphere oscillates in the vertical direction, we have to take much deep water depth such as $h/a=25$ (see figure \ref{f7}) to  reach the infinite water depth  results of Hulme \cite{Hu1982}.

\section{Discussion and conclusion}
To solve a wave body motion problem through a Green function method, it is necessary to evaluate the Green function $G$ and its gradient $\nabla G$. It is known that it is  troublesome in the calculation of $G$ and it is even much worse for the approximation of $\nabla G$.

For the free surface Green function  with respect to a  free surface oscillating source in a fluid of  infinite  depth, the function $G$ is expressed as
 the sum  of the Rankine simple Green function $1/r$, its image $1/r_1$ with regarding to the average free water surface and a singular wave integral. Similarly, the finite depth Green function is the sum of $1/r$ and its image $1/r_0$ with respect the fluid bottom $z=-h$ and a singular wave integral $G_1$ (see John \cite{Jon}):
 \bbe G= \frac1r+\frac1{r_0}+G_1.
 \bee
 Under this  formulation, the integrands of $G_1$ and $\nabla G_1$  on  the free surface  have the asymptotic behaviours (as $k\to \infty$)
  \bbe O(\frac{ \cos (kR - \frac \pi4)}{\sqrt{kR}}) \mbox{ and }  O( \sqrt{\frac kR}\cos (kR - \frac \pi4)),
  \bee
  respectively, as given in  (\ref{new11}) and (\ref{new33})).

 The principal difference between the infinite depth and finite depth Green functions is that the former has the single free surface mirror $z=0$ while the latter has the additional   water bed mirror  $z=-h$ in relation to  the image method. Due to the reflection in between the two mirrors,  for a free surface source point $\pp=(\xi,\eta,\zeta)$ in the fluid domain $-h<z<0$, we need the following  five   images of $\pp$:
 \be
 \pp_0&=& (\xi,\eta,-\zeta-2h), \mbox{ the image of $\pp$ with respect to the mirror $z=-h$,}
 \\
 \pp_1&=& (\xi,\eta,-\zeta), \mbox{ the image of $\pp$ with respect to the mirror $z=0$,}
 \\
 \pp_2&=& (\xi,\eta,\zeta-2h), \mbox{ the image of $\pp_1$ with respect to the mirror $z=-h$,}
 \\
 \pp_3&=& (\xi,\eta,\zeta+2h), \mbox{ the image of $\pp_0$ with respect to the mirror $z=0$,}
 \\
 \pp_4&=& (\xi,\eta,-\zeta-4h), \mbox{the image of $\pp_3$ with respect to the mirror $z=-h$.}
 \ee

 Therefore, in the present formulation, the Green function with respect to a field point $\q=(x,y,z)$ is expressed as
 \bbe G= \frac1{|\q-\pp|}+ \sum_{n=0}^4 \frac1{|\q-\pp_n|}+K.\bee
 Here $K$ is approximated by  a  wave integral, of which the integral is smooth and decay rapidly in the infinite frequency integral domain. This also gives rise to the new formulation of the vertical and horizontal  derivatives, for $r=|\q-\pp|$ and $r_n = |\q-\pp_n|$,
 \begin{align}
\p_\zeta G&= \p_\zeta(\frac1r+\frac1{r_0}+\frac1{r_1}+\frac1{r_2}+\frac1{r_3}+\frac1{r_4})+2\nu(\frac1{r_1}-\frac1{r_2}+\frac1{r_3}-\frac1{r_4})+K_1,
\\
 \p_R G &= \partial_R \left(\frac1{r}+\frac1{r_0}+\frac1{r_1}+\frac1{r_2}+\frac1{r_3}+\frac1{r_4}\right)+K_2\nonumber \\
&+ \frac{2\nu R}{r_1(r_1\!+\!|z\!+\!\zeta|)}\!+\!\frac{2\nu R}{r_2(r_2\!+\!|z\!-\!\zeta\!-\!2h|)}\!+\!\frac{2\nu R}{r_3(r_3\!+\!|\zeta\!-\!z\!-\!2h|)}\!+\!\frac{2\nu R}{r_4(r_4\!+\!|z\!+\!\zeta\!+\!4h|)},
\end{align}
with $K_1$ and $K_2$ the wave integrals defined respectively by (\ref{aG2}) and (\ref{aG3}). As illustrated in (\ref{nn1}) and (\ref{nn2}), the integrands of the wave  integrals $K$, $K_1$ and $K_2$ have the same asymptotic behaviours $O(\frac1{k^{3/2}})$   when both the field point $\q$ and the source point $\pp$ are on the free surface. Thus the divergence for the wave integrals from John's formula becomes   the convergence for  the modified wave integrals. This together with the use of the artificial parameter $\mu$, direct integral of  $K$, $K_1$ and $K_2$ can be performed on the straight line $0<k<\infty$.

  The Green function formula  (\ref{G1})  involves explicitly  the Rankine source potential $1/r$, the Rankine image source potential $1/r_0$ and the wave integral $G_1$. The integral is divergent at the infinity because the  other four Rankine image source potentials $1/r_1,...,1/r_4$ are contained  implicitly in $G_1$. We therefore  separate them  from $G_1$ to form $K$, which becomes convergent at the infinity.

     With the presence of the solid mirror $z=-h$ and the free surface mirror $z=0$, the  source $\pp$ in the fluid domain  has infinitely many  images due to continued reflections in between the mirrors. However, only the five images are useful in the new wave integral formulation.  If a Rankine  source is in between two parallel solid  mirrors without the  free surface effect, the Green function of this problem can be presented in a form involving  infinitely  many source  images (see \cite{1991}). This is  different to the present problem  due to the absence of free surface effect and thus  wave integral in \cite{1991}.


 Validation of the present method  is provided through comparison  between the present method results and
 the results from  John \cite{Jon} series together with the added mass and damping coefficients results  of   Linton \cite{Linton1991} and Wang \cite{W1986}  on submerged oscillatory sphere
 and Hulme \cite{Hu1982} on oscillatory floating hemisphere  in waves. 

 This research is motivated by the  direct integration approach of the wave integral  \cite{Chen2015} on double wave integral and \cite{Chen2019} on single wave integral approximations for the three-dimensional infinite depth  Green function with application to  an  oscillatory body motion in waves. This study is originated from  \cite{Chen2012} on two-dimensional vortex Green function method for  a  travelling body in  fluid.

 Nowadays,  numerical computation of a linear hydrodynamic problem  is no longer a time consuming job due to  the popularization of high  computing capacity computers. Nevertheless, the Green function evaluation due to the presence of an irregular wave integral is still known to be troublesome  and sophisticated mathematical approximation theories  are supposed to be employed to attack the evaluation \cite{N1985,Nob82,WuNob17,HessWilcox69,PeterM}. The present investigation however shows that the singularity is removable and an accurate evaluation of the Green function is obtainable  by an elementary integration  in a straightforward manner.

\appendix
\section{Derivation of John's Green function }
For the completion of the analysis, we follow John \cite{Jon} to show the  derivation of the Green function
$$G=\frac1r+\frac1{r_0}+G_1.$$

The use  of    the  Hankel transformation (\ref{G3})  produces
\be
{\mathcal H}(\frac1r+\frac1{r_0})&=& \frac1k\left(\e^{-k|z-\zeta|}
 +\e^{-k|z+\zeta+2h|}\right).
 \ee
Then  applying   the Hankel transformation  to the Laplace equation  (\ref{con1}) and employing  the bottom boundary condition (\ref{con3}), we have
\bbe
\H  (G_1 )=\frac1k A_0(k) \cosh k(z+h)\label{GG1}
\bee
for a function $A_0$.
Therefore  applying again the Hankel transformation  to the free surface condition (\ref{con2}), we have,  for $z$ close to $0$,
\bbe
0&\approx & (\p_z-\nu)\H (G_1) +\p_z{\mathcal H}(\frac1r+\frac1{r_0})-\nu {\mathcal H}(\frac1r+\frac1{r_0})\nonumber
\\
&=&\frac1k A_0(k) [k\sinh k(z+h)-\nu \cosh k(z+h)]- \frac{k+\nu}k\left(\e^{-k(z-\zeta)}
 +\e^{-k(z+\zeta+2h)}\right).
\bee
This implies that
\be
&&A_0(k)=\frac{2(\nu +k)\e^{-kh}\cosh k(\zeta+h)}{k\sinh kh-\nu \cosh kh }.
\ee
Therefore the desired Green function is obtained by rewriting  (\ref{GG1}) as
\be
&&  G_1=H^{-1}(\frac1k A_0(k) \cosh k(z+h)),
\ee
or the desired Green function
\bbe
 G&=&\frac1r+\frac1{r_0}+\int_L\frac{2(\nu +k)\e^{-kh}\cosh k(\zeta+h)\cosh k(z+h)}{k\sinh kh-\nu \cosh kh }J_0(kR)dk. \label{G4}
\bee
The integral pass $L$ passing beneath  the pole is determined by the asymptotic behaviour (\ref{con4}).

\

\noindent \textbf{Acknowledgement.} This work was partially  supported by NSFC of China (11571240).

\


\begin{thebibliography}{99}


\bibitem{Frank1967} W. Frank,  Oscillation of cylinders in or below the free surface of deep fluids,  Report  2375,  Naval Ship Research  Development Center, Bethesda, MD, 1967.

\bibitem{Lee1989}  C.H. Lee,   P.D. Sclavounos, Removing the irregular frequencies from integral equations in wave-body interactions, J. Fluid Mech. 207 (1989) 393-418.

\bibitem{LeeNew} C.H. Lee,  J.N. Newman, Computation of wave effects using the panel method, In: S.K. Chakrabart (Ed.), Numerical Models in Fluid-Structure Interaction, WIT Press, Southampton, 2004.

\bibitem{Chak} S.K Chakrabarti, Application and verification of deep water Green function for water waves, J. Ship Res. 45 (2001) 187-196.

\bibitem{LiangWuNob18} H. Liang, H. Wu, F. Noblesse. Validation of a global approximation to the Green function of diffraction radiation in deep water, Appl. Ocean Res. 74 (2018) 80-86.

\bibitem{N1985} J.N.  Newman,   Algorithms for the free-surface Green functions, J. Engng. Math. 19 (1985)  57-67.

\bibitem{Nob82} F. Noblesse, The Green function in the theory of radiation and diffraction of regular water waves by a body, J. Engng. Math. 16 (1982) 137-169.


\bibitem{PonBob94} B. Ponizy, F. Noblesse, M. Ba, M. Guilbaud, Numerical evaluation of free-surface Green function, J. Ship Res. 38 (1994) 193-202.


 \bibitem{TelNob86} J.G. Telste, F. Noblesse, Numerical evaluation of the Green function of water-wave radiation and diffraction, J. Ship Res. 30 (1986) 69-84.

 \bibitem{WuNob17} H. Wu, C. Zhang, Y. Zhu, W. Li, D. Wan, F. Noblesse, A global approximation to the Green function for diffraction radiation of water waves, European J. Mech. / B Fluids 65 (2017) 54-64.

\bibitem{Jon} F. John,  On the motion of floating bodies II. Simple harmonic motions, Communs. Pure Appl. Math. 3 (1950) 45-101.


\bibitem{Linton1999} C. M. Linton,   Rapidly convergent representations for Green functions for Laplace's equation, Proc. R. Soc. A  455 (1999) 1767-1797.


\bibitem{Liu} Y. Liu, H. Iwashita,  C. Hu, A calculation method for finite depth free-surface green function, Int. J. Nav. Archit. Ocean Eng. 7 (2015) 375-389.

\bibitem{Pid} M.K. Pidcock, The calculation of Green functions in three dimensional hydrodynamic gravity wave problems, Int. J. Numer. Meth. Fluids  5(1985) 891-909.

\bibitem{Have1955} T. Havelock,  Waves due to a floating hemi-sphere making periodic heaving oscillations, Proc. R. Soc. Lond. A 231 (1955) 1-7.

\bibitem{Hu1982} A. Hulme, The wave forces acting on a floating hemisphere undergoing forced periodic oscillations, J. Fluid Mech. 121 (1982) 443-463.

\bibitem{U1949} F. Ursell,  On the heaving motion of a circular cylinder on the surface of a fluid, Quart. J. Mech Appl. Math. 2 (1949) 218-231.

\bibitem{Fa} C. Farell, On the wave resistance of a submerged spheroid, J. Ship Res. 17 (1973) 1-11.

\bibitem{Cha2013}  I.K. Chatjigeorgiou, The analytic solution for hydrodynamic diffraction by submerged prolate spheroids in infinite water depth, J. Engng. Math. 81 (2013) 47-65.

\bibitem{W1986} S. Wang,  Motions of a spherical submarine in waves, Ocean Engng. 13 (1986) 249-271.

\bibitem{WuTaylor87} G.X. Wu,  R. Eatock Taylor, The exciting force on a submerged spheroid in regular waves, J. Fluid Mech. 182 (1987)  411-426.

\bibitem{Linton1991}C. M. Linton, Radiation and diffraction of water waves by a submerged sphere in finite depth, Ocean Engng. 18 (1991) 61-74.


\bibitem{Cao} Y. Cao, W. Schultz, R. Beck,  Three-dimensional desingularized boundary integral methods for potential problems, Int. J. Numer. Meth. Fluids 12 (1991) 785-803.

\bibitem{Dawson}  C.W. Dawson,  A practical computer method for solving ship wave problems, In {Proceedings of 2nd International Conference on
Numerical Ship Hydrodynamics}, University of California, Berkeley,  30-38, 1977.

\bibitem{FengChen1} A. Feng, Z.M. Chen, W.G. Price, A Rankine source computation for three dimensional wave-body interactions adopting a nonlinear body boundary condition, Appl. Ocean Res. 47 (2014) 313-321.

\bibitem{FengChen2} A. Feng, Z.M. Chen, W.G. Price, A continuous desingularized source distributi2n method describing wave-body interactions of a large amplitude oscillatory body, J. Offshore Mech. Arctic Engng. 137 (2015), 021302.

\bibitem{Man}  D.A. Mantzaris,  A Rankine panel method as a tool for the hydrodynamic design of complex marine vehicles, PhD thesis, MIT, 1998.

\bibitem{Y1981} R.W. Yeung,   Added mass and damping of a vertical cylinder in finite depth waters, Appl. Ocean Res. 3 (1981) 119-133.

 \bibitem{Chen2019} Z.M. Chen, Straightforward integration for free surface Green function and body
wave motions, European J. Mech. / B Fluids 74(2019) 10-18.

\bibitem{WehausenLaitone} J.V. Wehausen,  E.V. Laitone,   Surface waves, In:  S. Flugge, C. Truesdell (Eds.), Fluid Dynamics III in Handbuch der Physik 9, Springer, Berlin,   446-778,  1960.

\bibitem{Abr} M. Abramowitz, I.A. Stegun, Handbook of Mathematical Functions with Formulas,
Graphs, and Mathematical Tables, Dover, New York, 1965.



\bibitem{Hav28} T. H. Havelock,  Wave resistance, Proc. R. Soc. Lond. A 118 (1928), 24-33.

\bibitem{Hav32}T. H. Havelock, The theory of wave resistance, Proc. R. Soc. Lond. A 138 (1932), 339-348.

\bibitem{2009} L. V. Lazauskas, Resistance, wave-making and
wave-decay of thin ships, with emphasis on the effects of viscosity, PhD Thesis, The University of Adelaide, 2009.

\bibitem{2002} B. Spivak, J.-M. Vanden-Broeck, T. Miloh, Free-surface wave damping due to viscosity and surfactants,  European J. Mech. / B Fluids 21 (2002) 207-224

\bibitem{1967} V.J. Monacella, On ignoring the singularity in the numerical evaluation of Cauchy Principal Value integrals, Hydromechanics Laboratory Research and Development  Report 2356, 1967.



\bibitem{Wa} Watson, G. N., A Treatise on the Theory of Bessel Functions. Cambridge University Press, 1944.


\bibitem{HS} J.L. Hess,  A.M.O. Smith,    Calculation of non-lifting potential flow
about arbitrary three-dimensional bodies, Report No. E.S. 40622, Douglas Aircraft Co., Inc.  Aircraft Division, Long Beach, California, 1962. 

\bibitem{HS2} J.L. Hess,  A.M.O. Smith,  Calculation of  potential flow about arbitrary  bodies, Prog. Aerospace Sci. 8 (1966) 1-138.


\bibitem{Newman1986} J.N. Newman,   Distributions of sources and normal dipoles over a quadrilateral panel, J. Engng. Math. 20 (1986) 113-126.

\bibitem{U1981} F. Ursell, Irregular frequencies and the motion of floating bodies, J. Fluid Mech. 105 (1981) 143-156.


\bibitem{Lee1996} C.H. Lee,  J.N. Newman, X. Zhu, An extended boundary integral equation method for the removal of irregular frequency effects, Int. J. Numer. Meth. Fluids 23 (1996) 637-660.




\bibitem{ZhuLee1994}  X. Zhu,   C.H. Lee, Removing the irregular frequencies  in wave-body interactions. The 9th International Workshop on Water Waves and Floating Bodies, Japan, 245-249, 1994.



\bibitem{1991} S.R. Breit, The potential of a Rankine source between parallel planes and in
a rectangular cylinder, J. Engng. Math. 25 (1991), 151-163.

\bibitem{Chen2015} Z.M. Chen,  Regular wave integral approach to numerical simulation of radiation and diffraction of surface waves, Wave Motion 52 (2015) 171-182.

\bibitem{Chen2012} Z.M. Chen,  A vortex based panel method for potential flow simulation around a hydrofoil, J. Fluids Struct. 28 (2012) 378-391.


\bibitem{HessWilcox69} J.L. Hess and D.C. Wilcox, Progress in the solution of the problem of a three-dimensional body oscillating in the presence of a free surface - Final technical report, McDonnell Douglas Company Rep. DAC 67647, 1969.
    \bibitem{PeterM} M.A. Peter, M.H. Meylan, The eigenfunction expansion of the infinite depth free surface Green function in three dimensions, Wave Motion 40 (2004) 1-11.





















\end{thebibliography}
\end{document}